\documentclass[twocolumn]{aastex631}

\usepackage{graphicx}
\usepackage{multirow}

\usepackage{outlines}
\usepackage{enumitem}
\setenumerate[1]{label=\Roman*.}
\setenumerate[2]{label=\Alph*.}
\setenumerate[3]{label=\roman*.}
\setenumerate[4]{label=\alph*.}

\usepackage{placeins}
\usepackage{amsmath}
\newcommand{\degree}{$^{\circ}$}
\newcommand{\kms}{km s$^{-1}$}

\newcommand{\logeps}[1]{$\log \epsilon$(#1)}
\newcommand{\feh}{[Fe/H]}
\newcommand{\xh}[1]{[#1/H]}
\newcommand{\xm}[1]{[#1/M]}
\newcommand{\xfe}[1]{[#1/Fe]}
\newcommand{\bprp}{G$_{\rm BP}-$G$_{\rm RP}$}
\newcommand{\rettwo}{Ret \textsc{II}}
\newcommand{\psctwo}{Psc \textsc{II}}
\newcommand{\teff}{T$_{\rm eff}$}
\newcommand{\logg}{$\log g$}
\newcommand{\ax}[1]{A$_{\rm #1}$}

\begin{document}

\title{GHOST Commissioning Science Results: \\ Identifying a new chemically peculiar star in Reticulum II}

\correspondingauthor{Christian R. Hayes}
\email{Christian.Hayes@nrc-cnrc.gc.ca}
\email{chrishayesastro@gmail.com}
\newcommand{\affilhaa}{\affiliation{NRC Herzberg Astronomy and Astrophysics Research Centre, 5071 West Saanich Road, Victoria, B.C., Canada, V9E 2E7}}

\newcommand{\affiluvic}{\affiliation{Department of Physics and Astronomy, University of Victoria, Victoria, BC V8W 3P2, Canada}}

\newcommand{\affilaao}{\affiliation{Australian Astronomical Optics, Macquarie University, 105 Delhi Rd, North Ryde NSW 2113, Australia}}

\newcommand{\affilanu}{\affiliation{Research School of Astronomy and Astrophysics, College of Science, Australian National University, Canberra 2611, Australia}}

\newcommand{\affilgems}{\affiliation{Gemini Observatory/NSF’s NOIRLab, Casilla 603, La Serena, Chile}}

\newcommand{\affilgemn}{\affiliation{Gemini Observatory/NSF’s NOIRLab, 670 North A’oh$\rm \bar{o}$k$\rm \bar{u}$ Place, Hilo, HI 96720, USA}}


\author[0000-0003-2969-2445]{Christian R. Hayes}
\affilhaa

\author{Kim A. Venn}
\affiluvic

\author{Fletcher Waller}
\affiluvic

\author{Jaclyn Jensen}
\affiluvic

\author{Alan W. McConnachie}
\affilhaa, \affiluvic

\author{John Pazder}
\affilhaa, \affiluvic

\author{Federico Sestito}
\affiluvic

\author{Andr\'e Anthony}
\affilhaa

\author{Gabriella Baker}
\affilaao

\author{John Bassett}
\affilgems

\author{Joao Bento}
\affilanu

\author{Gregory Burley}
\affilhaa

\author{Jurek Brzeski}
\affilaao

\author{Scott Case}
\affilaao

\author{Edward Chapin}
\affilhaa

\author{Timothy Chin}
\affilaao

\author{Eric Chisholm}
\affilhaa

\author{Vladimir Churilov}
\affilaao

\author{Adam Densmore}
\affilhaa

\author{Ruben Diaz}
\affilgems

\author{Jennifer Dunn}
\affilhaa

\author{Michael Edgar}
\affiliation{Anglo Australian Observatory}

\author{Tony Farrell}
\affilaao

\author{Veronica Firpo}
\affilgems

\author{Joeleff Fitzsimmons}
\affilhaa

\author{Juan Font-Serra}
\affilgems

\author{Javier Fuentes}
\affilgems

\author{Colin Ganton}
\affilhaa

\author{Manuel Gomez-Jimenez}
\affilgems

\author{Tim Hardy}
\affilhaa

\author{David Henderson}
\affilgemn

\author{Alexis Hill}
\affilhaa

\author{Brian Hoff}
\affilhaa

\author{Michael Ireland}
\affilanu
\affilaao

\author{Venu Kalari}
\affilgems

\author{Neal Kelly}
\affilhaa

\author{Urs Klauser}
\affilaao

\author{Yuriy Kondrat}
\affilaao

\author{Kathleen Labrie}
\affilgemn

\author{Sam Lambert}
\affilhaa

\author{Lance Luvaul}
\affilanu

\author{Jon Lawrence}
\affilaao

\author{Jordan Lothrop}
\affilhaa

\author{G. Scott Macdonald}
\affilhaa

\author{Slavko Mali}
\affilaao

\author{Steve Margheim}
\affiliation{Rubin Observatory/NSF’s NOIRLab, Casilla 603, La Serena, Chile}

\author{Richard McDermid}
\affilaao

\author{Helen McGregor}
\affilaao

\author{Bryan Miller}
\affilgems

\author{Felipe Miranda}
\affilhaa

\author{Rolf Muller}
\affilaao

\author{Jon Nielsen}
\affilanu

\author{Ryan Norbury}
\affilhaa

\author{Oliver Oberdorf}
\affilgemn

\author{Naveen Pai}
\affilaao

\author{Gabriel Perez}
\affilgems

\author{Pablo Prado}
\affilgems

\author{Ian Price}
\affilanu

\author{Carlos Quiroz}
\affilgems

\author{Vladimir Reshetov}
\affilhaa

\author[0000-0001-5528-7801]{Gordon Robertson}
\affilaao

\author{Roque Ruiz-Carmona}
\affilgems

\author{Ricardo Salinas}
\affilgems

\author{Kim M. Sebo}
\affilanu

\author{Andrew Sheinis}
\affiliation{Canada-France-Hawaii-Telescope, Kamuela, HI, United States}

\author{Matthew Shetrone}
\affiliation{University of California Observatories, University of California Santa Cruz, Santa Cruz, CA, 95064}

\author{Keith Shortridge}
\affilaao

\author{Katherine Silversides}
\affilhaa

\author{Karleyne Silva}
\affilgems

\author{Chris Simpson}
\affilgemn

\author{Greg Smith}
\affilaao

\author{Kei Szeto}
\affilhaa

\author{Julia Tims}
\affilaao

\author{Eduardo Toro}
\affilgems

\author{Cristian Urrutia}
\affilgems

\author{Sudharshan Venkatesan}
\affilaao

\author{Lewis Waller}
\affilaao

\author{Ivan Wevers}
\affilhaa

\author{Ramunas Wierzbicki}
\affilhaa

\author{Marc White}
\affilanu

\author{Peter Young}
\affilanu

\author{Ross Zhelem}
\affilaao

\begin{abstract}

The Gemini High-resolution Optical SpecTrograph (GHOST) is the newest high resolution spectrograph to be developed for a large aperture telescope, recently deployed and commissioned at the Gemini-South telescope.  In this paper, we present the first science results from the GHOST spectrograph taking during its commissioning runs.  We have observed the bright metal-poor benchmark star HD 122563, along with two stars in the ultra faint dwarf galaxy, \rettwo{}, one of which was previously identified as a candidate member, but did not have a previous detailed chemical abundance analysis.  This star (GDR3 0928) is found to be a bona fide member of \rettwo{}, and from a spectral synthesis analysis, it is also revealed to be a CEMP-r star, with significant enhancements in the several light elements (C, N, O, Na, Mg, and Si), in addition to featuring an r-process enhancement like many other \rettwo{} stars. The light-element enhancements in this star resemble the abundance patterns seen in the CEMP-no stars of other ultra faint dwarf galaxies, and are thought to have been produced by an independent source from the r-process.  These unusual abundance patterns are thought to be produced by faint supernovae, which may be produced by some of the earliest generations of stars.

\end{abstract}

\keywords{Dwarf galaxies $-$ Chemically peculiar stars $-$ chemical abundances $-$ High resolution spectroscopy $-$ Observational astronomy}

\section{Introduction}

Reticulum \textsc{II} (\rettwo{}) is one of the closest ultra faint dwarf galaxies (UFD; M$_{\rm V} = -2.7$) to the Milky Way, discovered by the Dark Energy Survey \citep[DES;][]{Diehl2014} in the southern hemisphere (RA$=53.77$\degree, DEC=$-54.05$\degree, J2000).  
Low resolution spectroscopy of 17$-$25 members \citep{Koposov2015, Walker2015, Simon2015} found that \rettwo{} has a low mean metallicity ([Fe/H]$=-2.6$\footnote{In this paper we refer to both the ``absolute'' abundance, \logeps{X} $\equiv 12 + \log_{10}({\rm N_{\rm X}/N_{\rm H}})$, and abundances in bracket notation relative to the sun, e.g., \xfe{X} $\equiv \log_{10}({\rm N_{\rm X}/N_{\rm Fe}})_{*} - \log_{10}({\rm N_{\rm X}/N_{\rm Fe}})_{\odot}$}) with a significant metallicity dispersion  ($\sigma$[Fe/H]$\sim0.5$). Furthermore, examination of the radial velocities showed a small velocity dispersion ($\sigma_{\rm V}$ $\sim 3.5$ \kms), resulting in a high mass-to-light ratio ($>450$ M$_{\odot}$/L$_{\odot}$), typical of the UFDs.

The first examination of the chemical abundances in \rettwo{} from high resolution spectra showed that 7 (of 9) red giants are highly enriched in r-process elements \citep{Ji2016Nature}, a pattern reinforced by more detailed analyses which showed the r-process abundance pattern matches the well-studied metal-poor halo stars
\citep{Ji2016, Roederer2016}.
In addition, the non-enriched stars are amongst the most metal-poor stars in \rettwo{}, leading to the conclusion that \rettwo{} was enriched by a single rare and prolific r-process event after the initial star formation event \citep{Ji2023,Simon2023}.

The discovery of a neutron star merger, GW170817 \citep{Abbott2017} with an associated optical and infrared radiation matching theoretical models for a kilonova event \citet{Kasen2017, Chornock2017} revealed a potential source for r-process enrichment.  Kilonova events are also predicted as a source for the r-process nucleosynthesis of heavy elements, similar to the patterns seen in the \rettwo{} stars \citep[e.g.,][]{Thielemann2017,Drout2017, Metzger2019, Cowan2021}. 

However, \citet{Ji2023} and \citet{Simon2023} have recently argued that because the \xh{Ba} abundances are roughly constant in the r-process enhanced \rettwo{} stars, the r-processed gas should have been well mixed into the interstellar medium (ISM) of \rettwo{} before the births of these stars.  Thus, a short SFH history of \rettwo{} would seem to require a prompt r-process source on timescales $\lesssim$ a few 100 Myr, whereas NS mergers require binary evolution that can take Gyrs \citep[e.g., GW170817;][]{Blanchard2017, Pian2017}.

A short timescale for r-process enrichment re-opens the discussion of the source of this enhancement in \rettwo{}.  It could be explained by a NS merger that happened on a very short timescale, or some other prompt r-process enrichment event.  Typical core collapse supernovae (CCSNe) are known prompt events, but face some difficulties in reproducing the r-process pattern seen in the Sun \citep[][and references therein]{ Thielemann2017}. 
Other, rarer events, such as magnetorotationally powered SN jets \citep{Winteler2012,Nishimura2015,Nishimura2017,Mosta2018}, collapsar disk winds \citep{Woosley1993,MacFadyen1999,Miller2020}, or common envelope jet SNe \citep{Grichener2019, Grichener2022}, are, therefore, also candidates for the prompt enrichment in \rettwo{}.  Better understanding the mixing in \rettwo{} may place further limits on these timescales, which requires measuring the chemical abundances of more stars in \rettwo{}

Measuring detailed chemical abundances requires high-resolution spectroscopy, and yet UFDs are distant and poorly populated, such that meaningful samples of UFD stars requires pushing to the faint limits of the most sensitive spectrographs on the largest telescopes.  The Gemini High-resolution Optical SpecTrograph \citep[GHOST;][]{ghost_ireland} is the newest "workhorse" high-resolution spectrograph to be developed for a large aperture telescope and has recently been deployed and commissioned on the Gemini-South telescope \citep{ghost_mcconnachie}.  Here we present the first science results from the GHOST spectrograph, taken during the commissioning of the instrument to show the quality of the spectra and demonstrate its science potential.  Two stars have been observed in the UFD \rettwo{}; one with no previous high resolution spectral analysis.

This paper is organized as follows: Section \ref{sec:data} describes the GHOST spectra, including the data reduction processes.   Section \ref{sec:analysis} describes our abundance analysis method, and Section \ref{sec:results} presents our chemical abundance results.  Section \ref{sec:discussion} provides a discussion of the chemical evolution of \rettwo{} incorporating our results from GHOST.

\section{Observations and Data Reduction}
\label{sec:data}

\subsection{The Gemini High Resolution Optical Spectrograph }

GHOST is the newest instrument for the Gemini Observatory, and it is the newest ``workhorse'' high resolution spectrograph on an 8-m class telescope. 
GHOST consists of a Cassegrain unit mounted on the Gemini-South telescope that contains the positioning arm system for two sets of micro-lens based Integral Field Units (IFUs) that patrol a 7.5' field of regard. A fiber system feeds the light into a bench-mounted spectrograph located in the pier-lab at the Observatory. The IFUs image slice a 1.2'' stellar object by a factor of 5 in width using 19 fibers in the ``high resolution'' mode (hereafter HR, $R \sim 75,000$ for a single object) and by a factor of 3 using 7 fibers in the ``standard resolution'' mode (hereafter SR, $R \sim 50,000$ for one or two objects in dual-mode). Complete wavelength coverage from approximately 3600 \AA{} to beyond 1 $\mu$m is obtained in a single exposure in either mode. A variety of on-chip binning options are available in both the spatial and spectral directions depending on the magnitude of the target and the desired spectral resolution.

An overview of GHOST is provided in \citet{Mcconnachie2022_nature} and \citet{ghost_ireland}. More details on the Cassegrain unit and fiber system can be found in \citet{Zhelem2018,Zhelem2020} and \citet{Churilov2018}. The reader is referred to \citet{Pazder2016} for information on the optical design of the bench spectrograph, and the optomechanics are summarised by \citet{ghost_pazder}. The precision radial velocity capabilities and data reduction pipeline are described in \citet{Ireland2016,Ireland2018} and \citet{Hayes2022}. A forthcoming publication will provide a complete overview of the scientific functionality and delivered science performance of GHOST using on-sky data.

Two commissioning runs for GHOST occurred between June 20 - 30, 2022 and September 12 -- 16, 2022. A summary of the first run is presented in \cite{ghost_mcconnachie}. During the runs, a variety of targets were observed in order to verify requirements and to perform on-sky tests of the functionality of GHOST. Where possible and aligned with overall commissioning goals, scientifically compelling targets were selected to demonstrate the scientific capabilities of GHOST to the commissioning team and the international community. The three stars 
analyzed and presented in this paper are the first results from these commissioning targets.

\subsection{Targeting and Observations}

\begin{figure*}
  \centering
  \includegraphics[scale=0.45,trim = 0.in 0.in 0.in 0.in, clip]{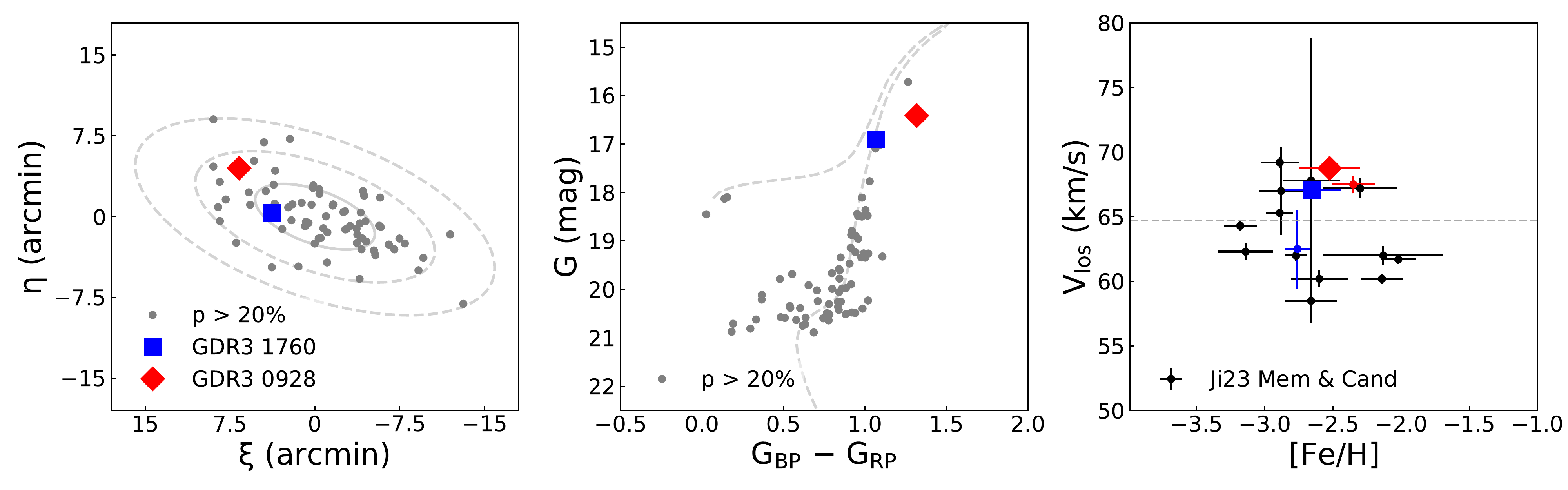}
  \caption{(Left) On sky distribution of stars with $> 20\%$ probability (gray points) for being \rettwo{} members from \citet{Jensen2023} over the one (solid), two, and three (both dashed lines) half-light radii ellipses (light gray lines), and highlighting GDR3 0928 (red diamond) and GDR3 1760 (blue square).  Positions are given relative to the \rettwo{} RA (03:35:42.1, h:m:s) and Dec (-54:02:57.0, d:m:s) (Middle) Gaia DR3 \bprp{} vs. G color-magnitude diagram of the probability $> 20\%$ stars with GDR3 0928 and GDR3 1760 highlighted as before.  (Right) Radial velocities and metallicities of \rettwo{} stars from \citet{Ji2023} and the two stars from this paper.  Stars from \citet{Ji2023} are shown as points (black), other than their values for GDR3 0928 and GDR3 1760 which have been colour-coded the same as for our measurements for ease of comparison.  The literature value \citep{McConnachie2012} for the systemic radial velocity of \rettwo{} is also marked (dashed gray line).}
  \label{fig:targeting}
\end{figure*}

Potential members of Milky Way dwarf galaxies have been identified from a maximum likelihood method developed by \citet{McVenn2020, McVenn2020DR3} and improved by \citet{Jensen2023}. Members are selected based on their spatial positions, colors, magnitudes, and proper motions using photometry and astrometry from {\it Gaia} DR3, and adopting a two component model for the distribution of stars in the satellite galaxies.  
This method has been shown to be efficient at identifying new members in both UFDs \citep{Waller2023} and classical dwarfs \citep{Sestito2023, Sestito2023b}.  
For \rettwo{}, targets with membership probabilities $>20\%$ from \citet{Jensen2023} are shown in Figure \ref{fig:targeting}.

All of the bright red giant branch (RGB) stars in \rettwo{} have been previously observed with high resolution spectroscopy with one exception, 
Gaia DR3 4732600514724860928 
(hereafter GDR3 0928).  This candidate is one of the brightest stars in \rettwo{}, with a $>99\%$ probability for membership, yet it only has a limited chemical abundance analysis in the literature \citep{Ji2023}. 
This star has been identified as a potential \rettwo{} member \citep{Massari2018, Ji2023} as it has a consistent metallicity and radial velocity with other stars in \rettwo{} \citep[Star 97 in][]{Ji2023}, but it has a redder color than the other \rettwo{} RGB stars, which can be seen in Figure \ref{fig:targeting}.  
The discrepancy between its red color and previously estimated low metallicity, \xh{Fe} $ = -2.35$, has precluded its identification as a bona fide member of \rettwo{}.  However, \citet{Ji2023} noted that this may also be due to a carbon enhancement.  Thus, we  selected GDR3~0928 for GHOST commissioning to test this hypothesis for its red color and measure its detailed chemistry.  

In addition to GDR3 0928, the previously known \rettwo{} member Gaia DR3 4732598457436901760 \citep[hereafter GDR3 1760, also referred to as DESJ033607-540235 by][]{Ji2016} was also observed with GHOST, in the standard resolution dual-object observing mode, allowing us to directly compare the spectra of this candidate member with a spectrum of a known member, taken under the exact same observing conditions.
The metal-poor benchmark RGB star, HD 122563, was also observed by GHOST in the standard resolution single-object observing mode, to provide a comparison star observed at a very high signal-to-noise ratio (SNR).
Positional and other basic targeting information for these stars is given in Table \ref{tab:targeting}.  
The total exposure times (split among three exposures) and additional specifications (binning in the order spectral x spatial) are given in Table \ref{tab:observing}, along with estimates of the final, order-combined SNR per pixel for each spectrum.

\begin{deluxetable*}{c c c c c c c c c c c}
\tablewidth{0pt}
\tablecolumns{11}
\tabletypesize{\scriptsize}
\tablecaption{Target Information and Radial Velocities \label{tab:targeting}}
\tablehead{\colhead{Name} & \colhead{{\it Gaia} DR3} & \colhead{RA (J2000)} & \colhead{Dec (J2000)} & \colhead{Gaia G} & \colhead{\bprp{}} & \colhead{E (B$-$V)} & \colhead{dist} & \colhead{$\sigma_{\rm dist}$} & \colhead{V$_{\rm los}$} & \colhead{$\sigma_{\rm V}$} \\ 
\colhead{} & \colhead{Source ID} & \colhead{(h:m:s)} & \colhead{(d:m:s)} & \colhead{(mag)} & \colhead{(mag)} & \colhead{(mag)} & \colhead{(kpc)} & \colhead{(kpc)} & \colhead{(\kms)} & \colhead{(\kms) }}
\startdata
HD 122563 & 3723554268436602240 & 14:02:31.84 & +09:41:09.9 & 5.87 & 1.22 & 0.0213 & 0.318 & 0.004 & -26.3 & 0.1 \\
GDR3 1760 & 4732598457436901760 & 03:36:07.76 & -54:02:35.5 & 16.90 & 1.07 & 0.0155 & 30.2 & 3.0 & 67.1 & 0.5 \\
GDR3 0928 & 4732600514724860928 & 03:36:27.68 & -53:58:26.2 & 16.41 & 1.31 & 0.0148 & 30.2 & 3.0 & 68.8 & 0.2 \\
\enddata
\end{deluxetable*}

\begin{deluxetable*}{c c c c c c c c}
\tablewidth{0pt}
\tablecolumns{8}
\tabletypesize{\scriptsize}
\tablecaption{Observing Information and SNRs \label{tab:observing}}
\tablehead{\colhead{Name} & \colhead{Date} & \colhead{Resolution} & \colhead{Single/Dual} & \colhead{Binning} & \colhead{Blue Exp.} & \colhead{Red Exp.} & \colhead{S/N} \\ \colhead{} & \colhead{(Y/M/D)} & \colhead{Mode} & \colhead{Target} & \colhead{(spec, spat)} & \colhead{Time (s)} & \colhead{Time (s)} & \colhead{4000/5200/6500/8500 \AA{}}}
\startdata
HD 122563 & 2022/06/28 & HR & Single & 1x1 & 180 & 30 & 270/650/420/480 \\
GDR3 1760 & \multirow{2}{*}{2022/09/14} & \multirow{2}{*}{SR} & \multirow{2}{*}{Dual} & \multirow{2}{*}{2x4} & \multirow{2}{*}{5400} & \multirow{2}{*}{5400} & 8/27/50/55 \\
GDR3 0928 & & & & & & & 7/30/60/70
\enddata
\end{deluxetable*}

\subsection{Data Reduction and Processing}

Spectra were reduced using the GHOST Data Reduction pipeline \citep[GHOSTDR;][]{Ireland2018, Hayes2022} through the DRAGONS framework \citep{dragons}, producing 1D order-combined sky-subtracted spectra for the blue and red camera separately.  For each of these 1D spectra, we perform a coarse normalization using a 4th order polynomial fit to the blue and red camera spectra independently.  We then use the {\texttt{Doppler}} code\footnote{\url{https://github.com/dnidever/doppler}} to cross correlate the spectra with template spectra of metal-poor stars synthesized with Turbospectrum v19.1.4 \citep{AlvarezPlez1998, Plez2012} and MARCS \citep{marcs} model atmospheres (following the synthesis details given in Section \ref{sec:analysis}) to calculate our radial velocities.  

Spectra were barycentric corrected before calculating the line-of-sight radial velocities (V$_{\rm los}$) in Table \ref{tab:targeting}.  We find the measured V$_{\rm los}$ for GDR3 0928 and GDR3 1760 are in good agreement with other stars in \rettwo{} from \citet{Ji2023}. The radial velocity for HD 122563 is also in good agreement with the result in the Gaia DR3 database (Gaia DR3 V$_{\rm los}$ = -26.13 \kms). 
For GDR3 0928, there is slight difference ($1.8\sigma$) between our value and that measured in \citet{Ji2023}.  GDR3 1760 has much larger differences between previous measurements and this work up to even 5 km/s, which is consistent with the observation by \citet{Ji2023} that this star is a radial velocity variable star, and likely in a binary system.  

\begin{figure*}
  \centering
  \includegraphics[scale=0.4,trim = 0.in 0.in 0.in 0.in, clip]{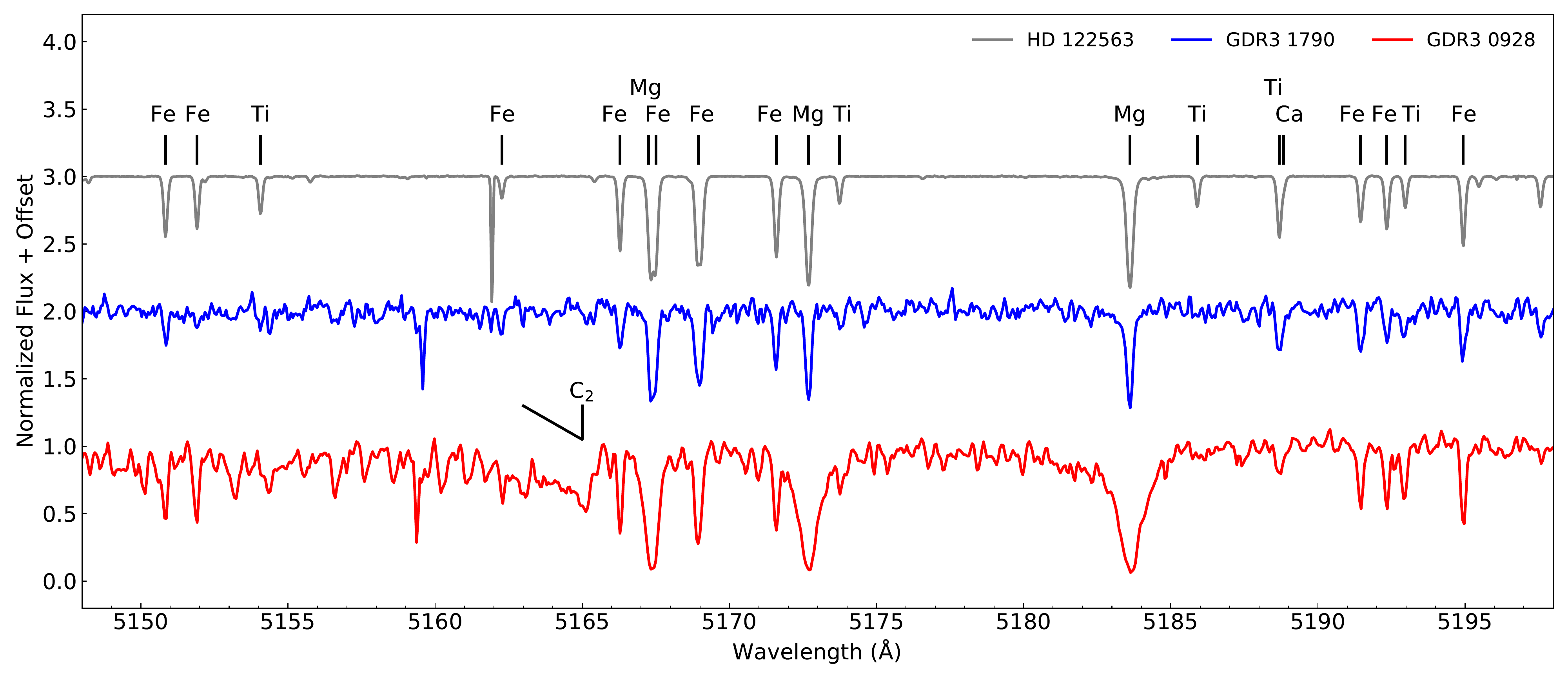}
  \caption{Spectra of HD 122563, GDR3 1760, and GDR3 0928, all with similar stellar parameters and metallicities (\teff{} $ = 4500-5000$ K and  \feh{} $= -2.8$ to $-2.5$) around the Mg b Triplet.  Notable is the 5165 \AA{} C$_2$ Swan band head and strong Mg b Triplet lines in GDR3 0928, while the Fe and Ti lines in this region are relatively comparable to HD 122563 and GDR3 1760.}
  \label{fig:spec_mgb}
\end{figure*}

After radial velocity correction, we use our spectral templates to identify continuum points in the observed spectra and fit a spline with iterative sigma clipping ($^{+10}_{-2} \sigma$) to produce a normalized spectrum.  The continuum normalized red and blue camera spectra were then stitched together with inverse variance weighting in the overlap spectral regions.
Figure \ref{fig:spec_mgb} shows an example of these reduced, normalized, and radial velocity corrected spectra around the \ion{Mg}{1} b Triplet.  We can easily see that GDR3 0928 has much broader \ion{Mg}{1} b Triplet lines and a prominent C$_2$ Swan band (e.g., the 5165 \AA{} bandhead).  The presence of C$_2$ absorption indicates that GDR 0928 is C-enhanced (confirmed through our abundance analysis in Section \ref{sec:analysis}). This confirms \citet{Ji2023}'s speculation that its red colour is due to carbon-enhancement despite its low \xh{Fe}.

The broad Mg b Triplet lines could either come from high surface gravity that broaden the lines, or such high Mg abundances that these lines are collisionally broadened. 
However, as GDR3 0928 has a measured parallax consistent with zero (0.005 $\pm$ 0.038 mas/yr), then it is inconsistent with being nearby dwarf star.  Thus, the strong Mg lines suggest that GDR3 is strongly enhanced in Mg, confirmed by our abundance analysis in Section~\ref{sec:analysis}.

\section{Analysis}
\label{sec:analysis}

\subsection{Initial Stellar Parameters}

Initial surface temperatures were determined from the Gaia calibrated infrared flux method \citep[IRFM][]{GHB2009, Mucciarelli2021} using the dust reddening maps from \citet{sfd_dust}\footnote{\citet{sfd_dust}
dust maps available at
\url{https://irsa.ipac.caltech.edu/applications/DUST/}. We adpoted the following conversion factors: \ax{V}/E(B$-$V) $= 3.1$, \ax{G}/\ax{V} $= 0.85926$, \ax{BP}/\ax{V} $= 1.06794$, \ax{RP}/\ax{V} $= 0.65199$ \citep{Marigo2008, Evans2018}}. 
The initial surface gravities were derived from the Stefan-Boltzmann law
using the IRFM temperatures, and the distance to HD 122563 from its {\it Gaia} DR3 parallax \citep{gaiadr3} and the distance to \rettwo{} from \citet{Koposov2015}.  
The surface gravity determinations also required; 1) a starting metallicity,  taken from \citet{Collet2018} for HD 122563, \citet{Ji2016} for GDR3 1760, and \citet{Ji2023} for GDR3 0928, and 
2) a bolometric correction, which we take from \citet{Andrae2018} assuming our targets are all giants.
The surface gravities and uncertainties are then determined with a Monte Carlo sampling of the uncertainties on the 
initial temperatures and distances, and using a flat mass prior appropriate for old RGB star \citep{Sestito2023}. We note that small changes in metallicity have a very small impact on the resulting temperatures and surface gravities.

Microturbulence velocities are from the relationship with surface gravity for giants from \citet{Mashonkina2017}. 
The C and $\alpha$-element abundances in the model atmospheres were initially set to the \xm{C} and \xm{$\alpha$}\footnote{Throughout this paper, when not referring to specific element, we use the following notations for different abundance representations.  For bulk atmospheric metallicities we refer to these with ``M,'' e.g., \xh{M}, to disambiguate these metallicities, since multiple elements are tied to these values and as opposed to \feh{} measured by Fe alone.  We use ``$\alpha$,'', ``r,'' and ``s'' to refer generically to $\alpha$-elements, r-process, and s-process elements respectively.  And finally we use the symbol ``X'' to refer to an individual, but unspecified element.} to the C and Mg abundances from \citet{Ji2016} for GDR3 1760, to 0.0 and 0.4 for HD 122563, and to +1 for GDR3 0928 given the strong C and Mg features observed in the spectrum.  These initial C and $\alpha$-element abundances were later altered to match the C and Mg abundances per star.

\begin{figure*}
  \centering
  \includegraphics[scale=0.4,trim = 0.in 0.in 0.in 0.in, clip]{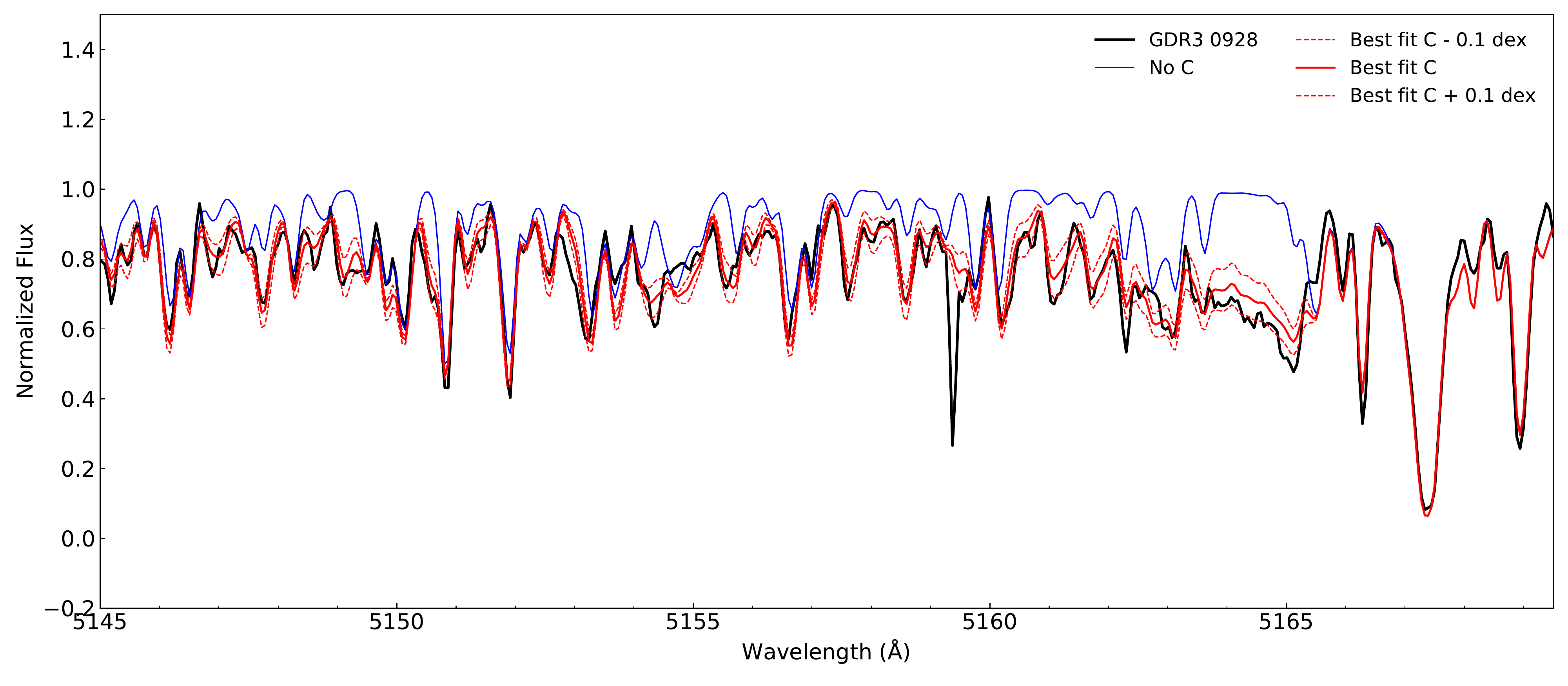}
  \caption{Spectrum of GDR3 0928 around the 5165\AA{} C$_2$ Swan band head compared to synthetic spectra with no carbon, our best fit carbon abundance $\log{\epsilon} = 7.95$, and 0.1 dex around our best fit carbon abundance.}
  \label{fig:cand_swanband}
\end{figure*}

\begin{figure*}
  \centering
  \includegraphics[scale=0.35,trim = 0.in 0.in 0.in 0.in, clip]{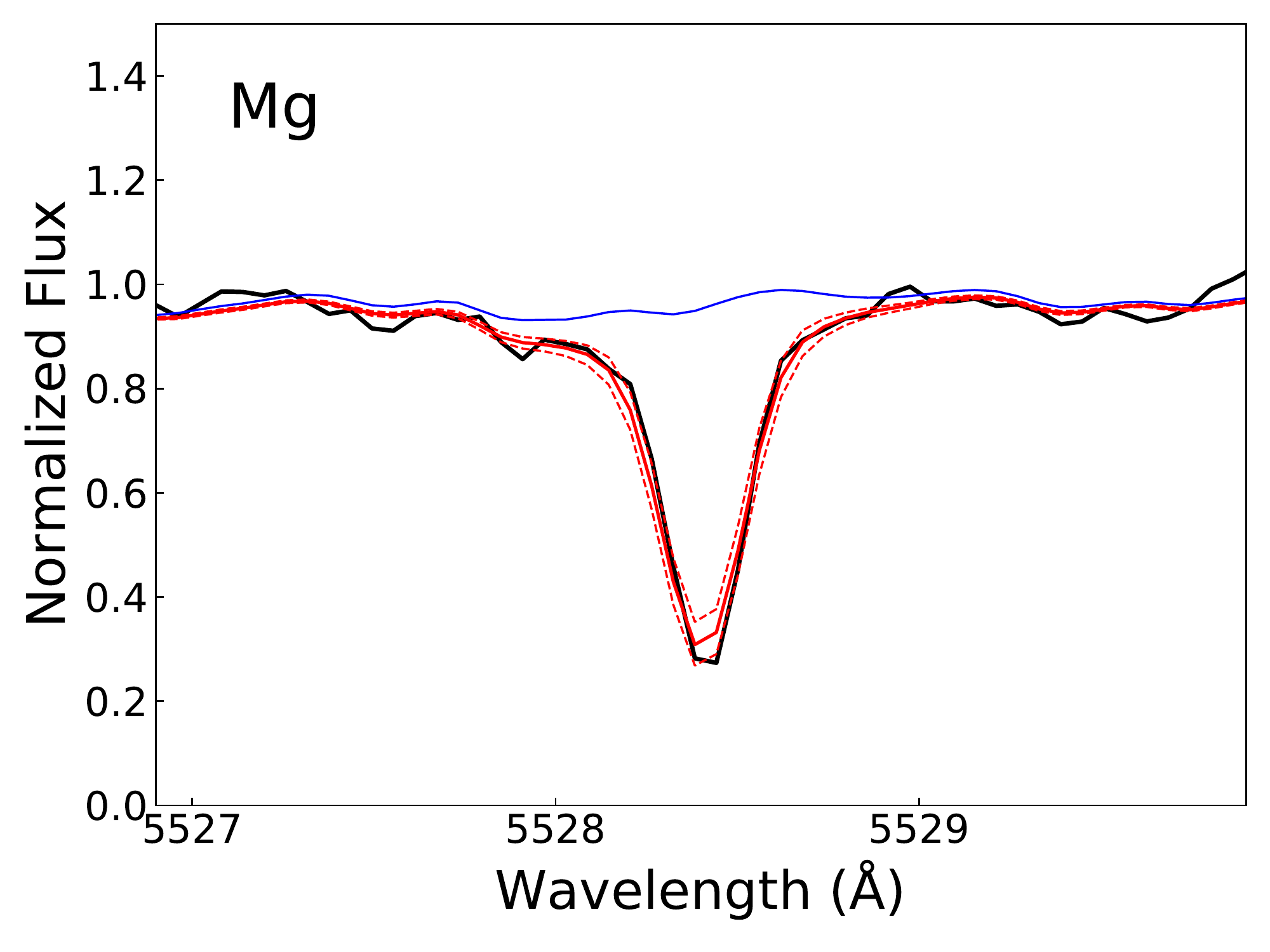}
  \includegraphics[scale=0.35,trim = 0.in 0.in 0.in 0.in, clip]{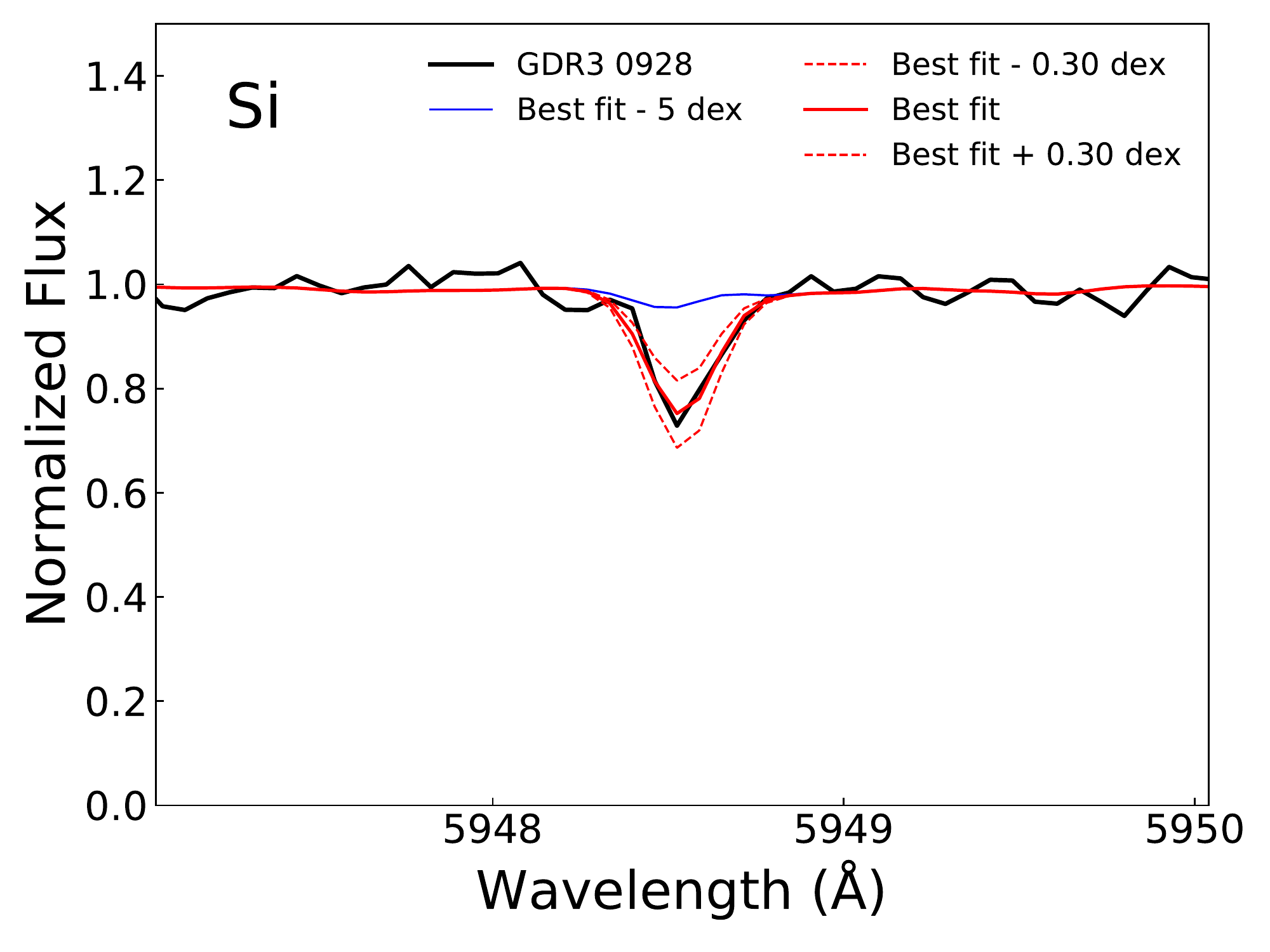}
  \includegraphics[scale=0.35,trim = 0.in 0.in 0.in 0.in, clip]{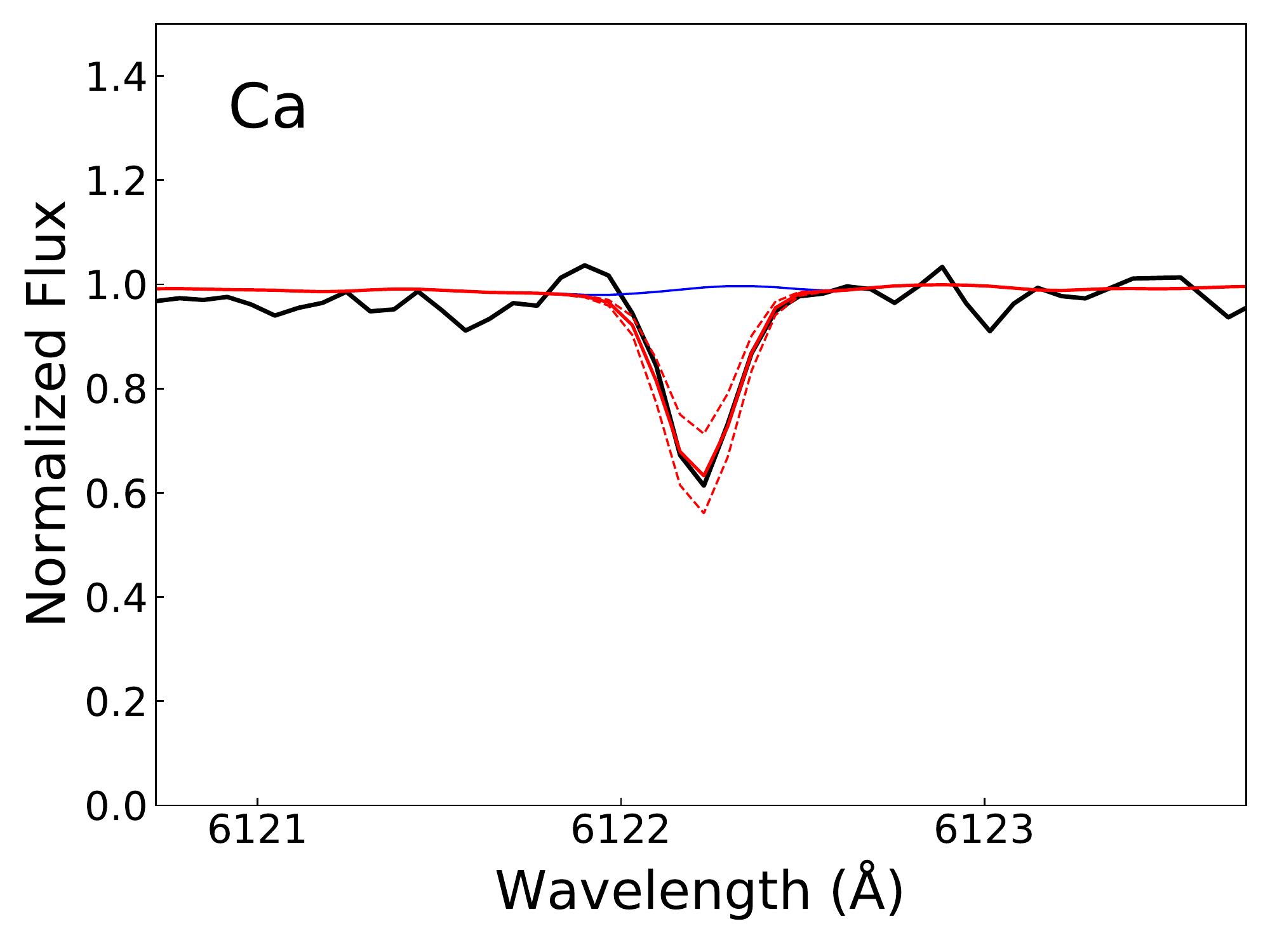}
  \includegraphics[scale=0.35,trim = 0.in 0.in 0.in 0.in, clip]{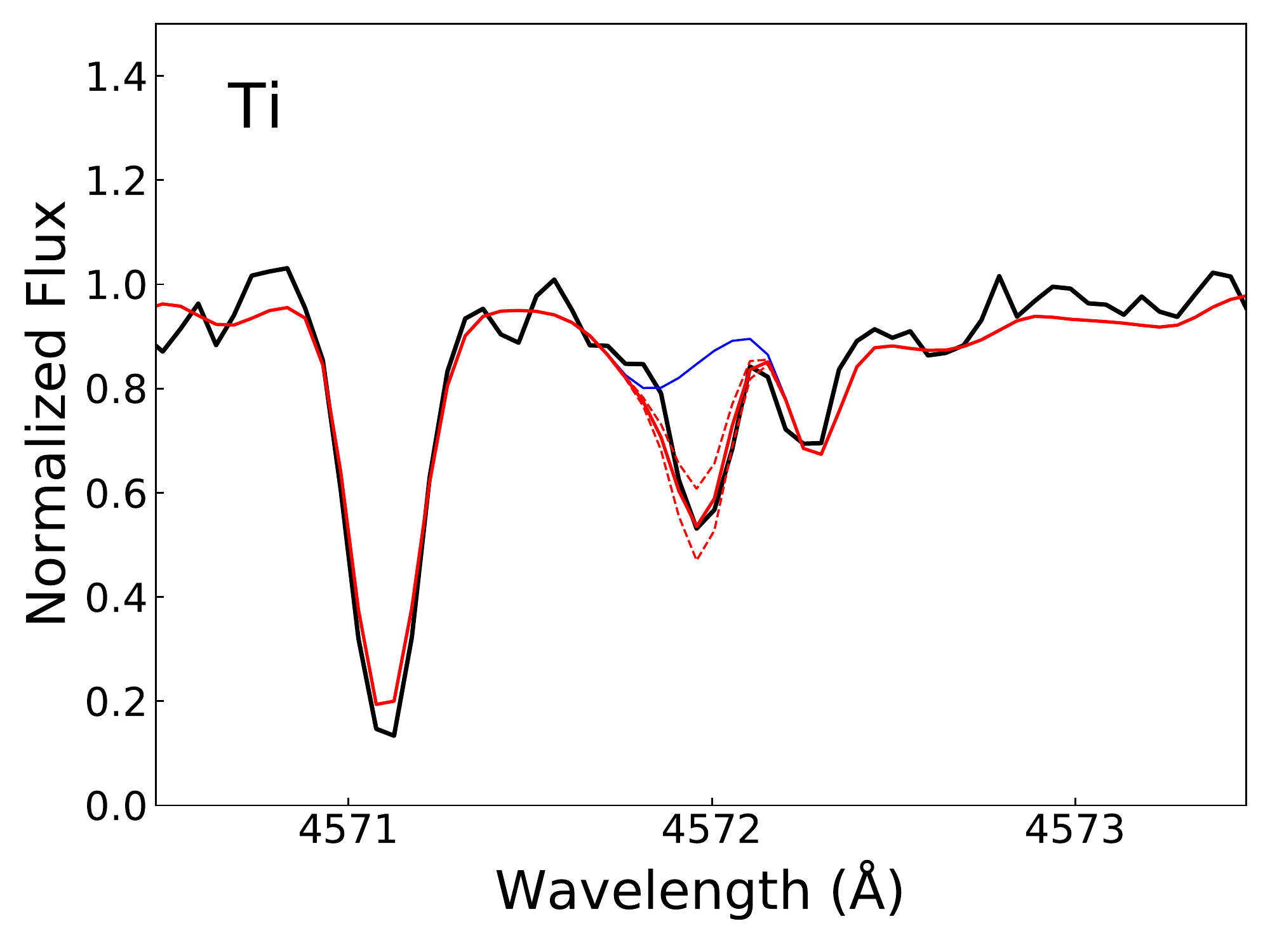}
  \includegraphics[scale=0.35,trim = 0.in 0.in 0.in 0.in, clip]{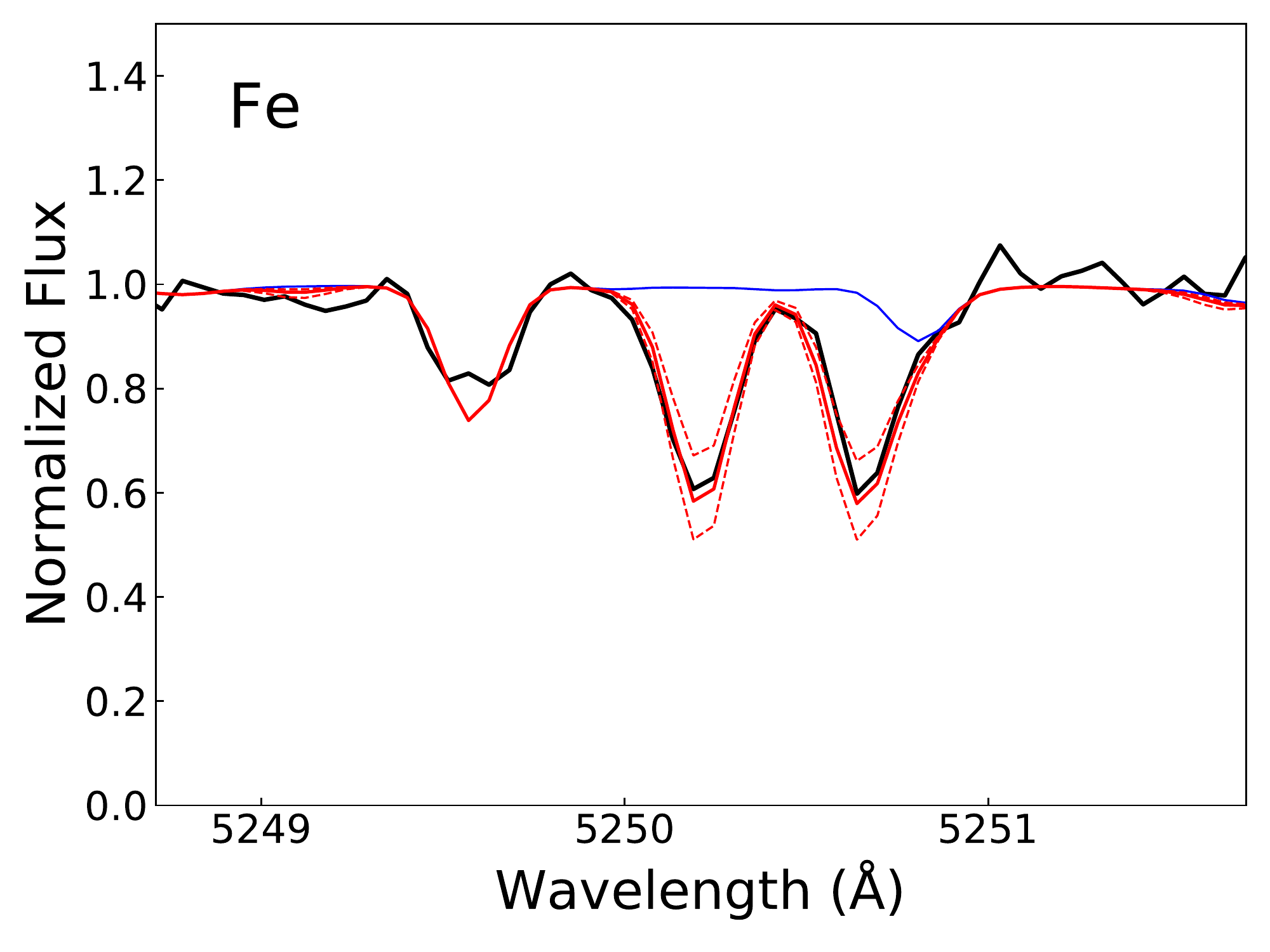}
  \includegraphics[scale=0.35,trim = 0.in 0.in 0.in 0.in, clip]{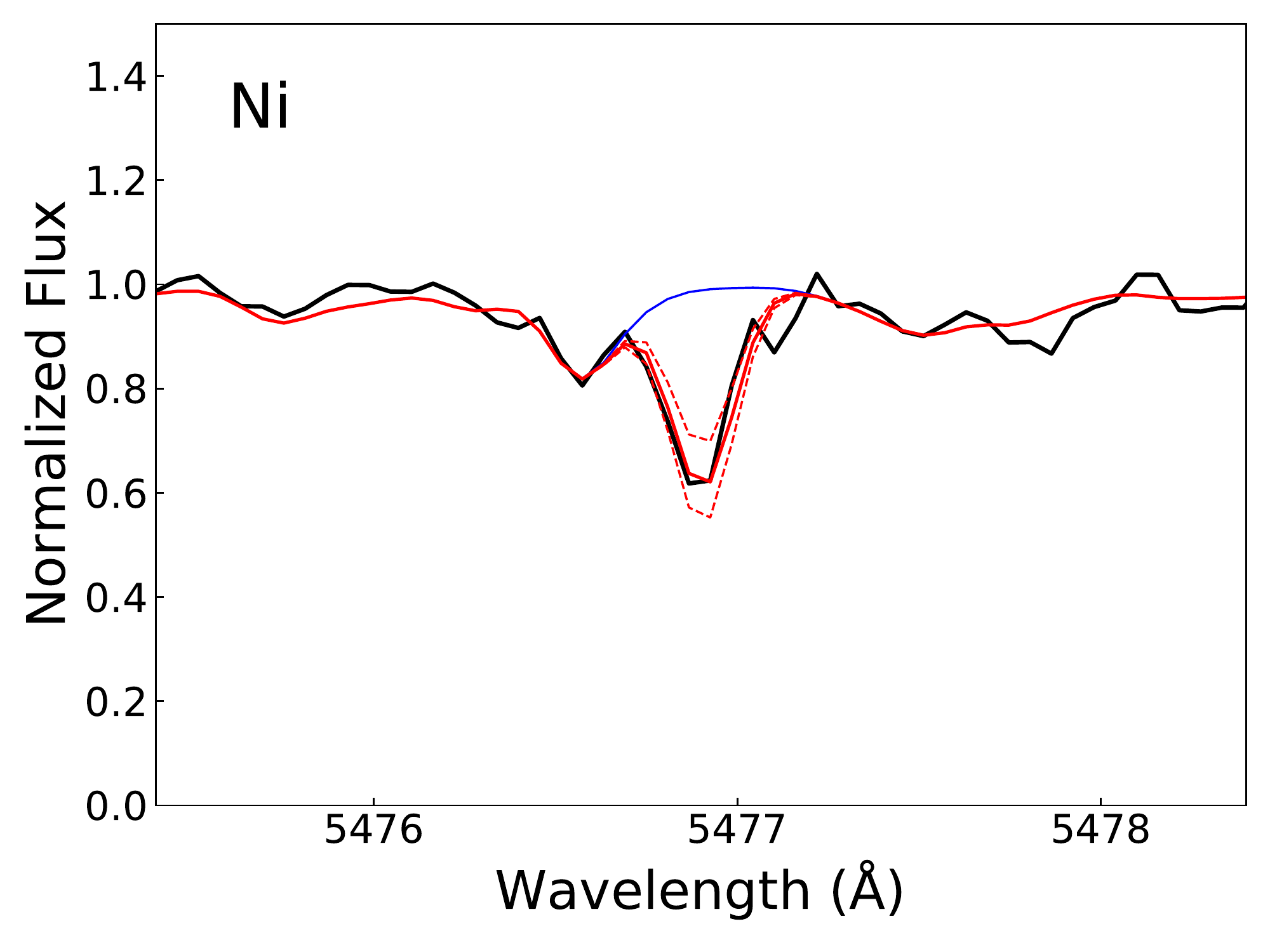}
  \includegraphics[scale=0.35,trim = 0.in 0.in 0.in 0.in, clip]{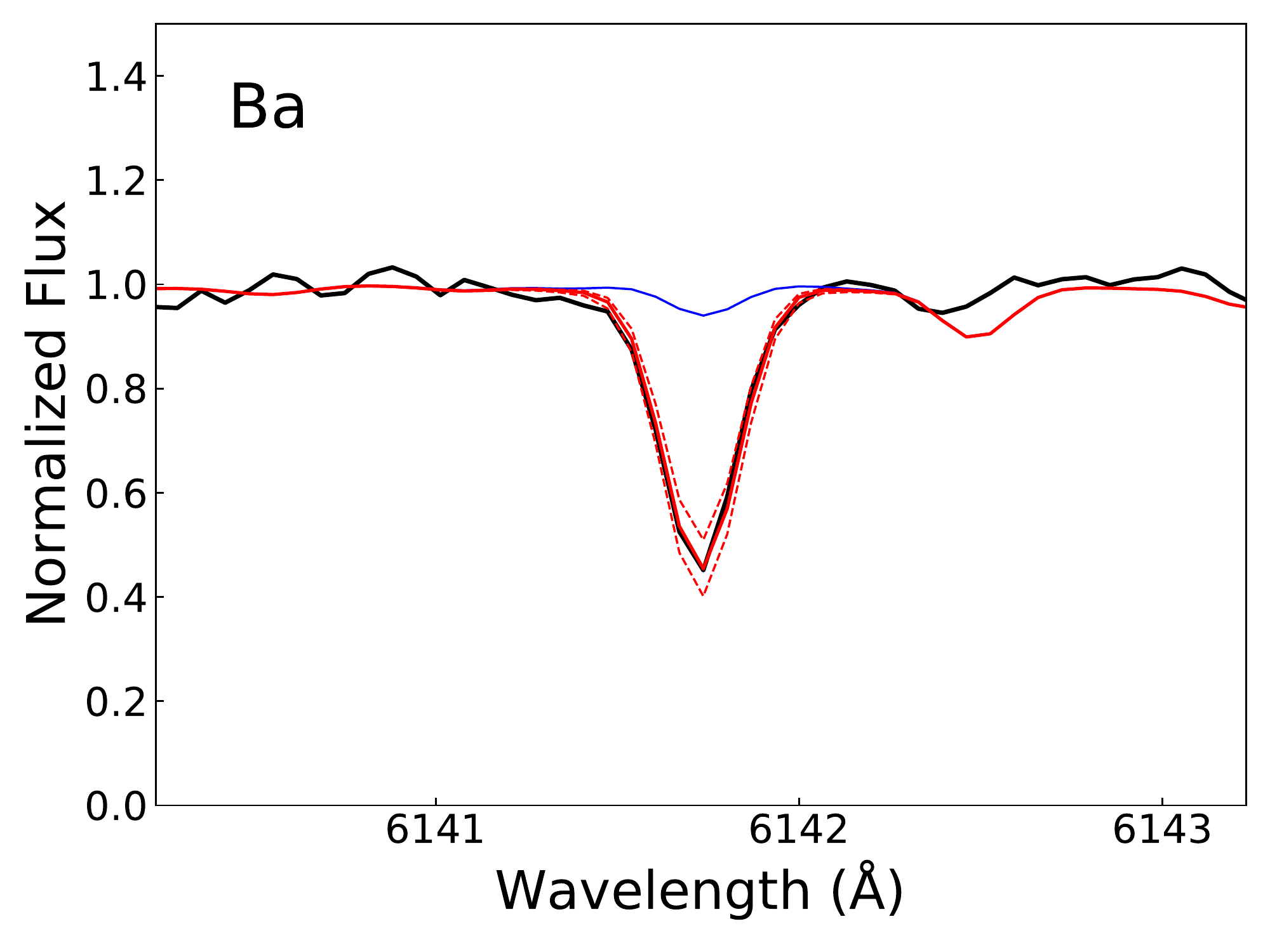}
  \includegraphics[scale=0.35,trim = 0.in 0.in 0.in 0.in, clip]{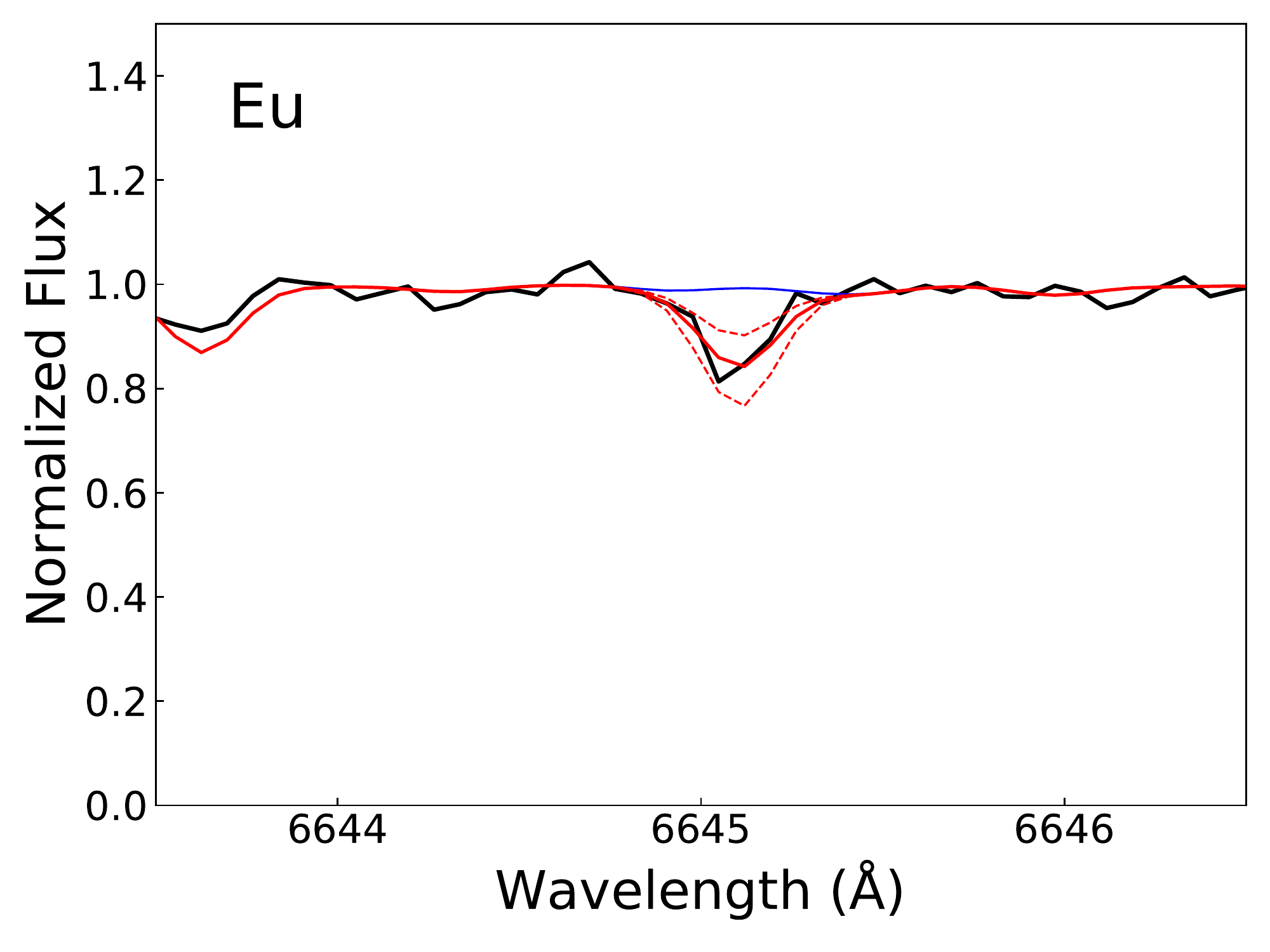}
  \caption{Spectrum of GDR3 0928 compared to synthetic spectra with with our best fit abundances for lines of different elements, 0.3 dex around the best fit for that line, and without that element.  All other elements in the syntheses are held fixed to the final abundance measured for that element.}
  \label{fig:cand_fits}
\end{figure*}

\begin{deluxetable*}{c c c c c c c c c c c}
\tablewidth{0pt}
\tablecolumns{11}
\tablecaption{Final Atmospheric Stellar Parameters \label{tab:params}}
\tablehead{\colhead{Name} & \colhead{\teff} & \colhead{$\sigma_{\rm Teff}$} & \colhead{\logg} & \colhead{$\sigma_{\rm logg}$} & \colhead{\xh{M}} & \colhead{$\sigma_{\rm \xh{M}}$} & \colhead{V$_{\rm micro}$} & \colhead{$\sigma_{\rm micro}$} & \colhead{\xm{C}} & \colhead{\xm{$\alpha$}} \\ \colhead{} & \colhead{(K)} & \colhead{(K)} & \colhead{} & \colhead{} & \colhead{} & \colhead{} & \colhead{(\kms)} & \colhead{(\kms)} & \colhead{} & \colhead{}}
\startdata
HD 122563 & 4642 & 88 & 1.26 & 0.07 & -2.83 & 0.15 & 2.1 & 0.1 & -0.3 & 0.6 \\
GDR3 1760 & 4986 & 108 & 1.95 & 0.12 & -2.63 & 0.21 & 1.9 & 0.1 & 0.3 & 0.3 \\
GDR3 0928 & 4507 & 80 & 1.87 & 0.2 & -1.3 & 0.2 & 1.8 & 0.1 & 1.0 & 1.0 
\enddata
\end{deluxetable*}

\subsection{Analysis Methodology}

In this work we determine chemical abundances using spectral synthesis.  Our synthetic stellar spectra are produced using the LTE spectral synthesis code Turbospectrum \citep[v19.1.4;][]{AlvarezPlez1998, Plez2012}, with 1D spherical radiative transfer.  The model atmospheres for this analysis are derived from the MARCS models \citep{marcs} in the APOGEE DR17 \citep{sdssdr17, Holtzman2023} grid, which includes models with a range of carbon and $\alpha$-element enhancements spanning [C/M] and [$\alpha$/M] from $-1$ to 1 in steps of 0.25 dex.  
We use the MARCS interpolator to interpolate the stellar parameters T$_{\rm eff}$, $\log g$, and \xh{M}, at the nearest \xm{C} and \xm{$\alpha$} points in the grid.

It is important to properly account for the impacts of stellar and instrumental broadening on stellar spectra. We convolve synthetic our synthetic spectra with a Gaussian broadening of 8 \kms\ (full-width half-max) for our high resolution GHOST spectrum of HD 122563  and 10 \kms\ for our standard resolution GHOST spectra of GDR3 0928 and GDR3 1760 in order to account for the combination of stellar and instrumental effects.  
These broadening parameters were verified by comparing the abundances of several strong, unsaturated Fe I lines, derived from equivalent width and minimum $\chi^2$ measurements, and confirming that there was a good agreement.

The Kurucz linelist\footnote{\url{http://kurucz.harvard.edu/linelists/ gfnew/gfallwn08oct17.dat}} \citep{Kurucz2017} was used as a starting point, with updated atomic line data compiled in \texttt{linemake}\footnote{\texttt{linemake} contains laboratory atomic data (transition probabilities, hyperfine and isotopic substructures) published by the Wisconsin Atomic Physics and the Old Dominion Molecular Physics groups. These lists and accompanying line list assembly software have been developed by C. Sneden and are curated by V. Placco at  \url{https://github.com/vmplacco/linemake}} \citep{linemake, Placco2021}, which included isotopic and hyperfine substructures (i.e., Sc II, Mn I,
Ba II, and Eu II).  In addition to this we add the following 
molecular features:  CH from \citet{Masseron2014}, MgH from \citep{Kurucz2017} and linelists provided in the Exomol project database \citep{exomol}:  C$_2$ \citep{Yurchenko2018, McKemmish2020}, CN \citep{Brooke2014, Syme2021}, NH \citep{Brooke2014nh, Brooke2015, Fernando2018, Bernath2020}, OH \citep{Brooke2016, Yousefi2018}, SiH \citep{Yurchenko2018sih}, CaH \citep{Shayesteh2004, Li2012, Alavi2017, Bernath2020}, and FeH \citep{Dulick2003, Bernath2020}.


Appendix \ref{app:spectral_analysis} describes our analysis methodology in more detail for completeness.  Briefly, abundances are measured for individual spectral features using a minimum $\chi^2$ fitting of synthetic spectra to each line of interest in small spectral windows.   Poorly fit or uncertain lines are rejected.  Blends are important when using this method (though we favor unblended lines in this analysis when possible and reject lines with poorly fit, strong blends).
Our approach to account for blending was to fit Fe first, followed by elements critical for blends (i.e., Mg, C, N Ti, Ca, etc.).   We iterate on these fits at least once to ensure that weak blends have been taken into account.

\subsection{Fine Tuning the Stellar Parameters}

Stellar parameters are often determined by reducing trends of Fe abundances with excitation potential (EP) and equivalent width (EW) to tune the \teff{}, microturbulent velocity respectively, and reducing the differences between the \ion{Fe}{1} and \ion{Fe}{2} abundances to determine the \logg.  However, there are known issues with this method, particularly for metal-poor giants.  Low excitation potential Fe lines are known to be overabundant when using the best interferometric measurements in 1D LTE, primarily due to unaccounted for 3D effects, and can bias traditional spectroscopic temperature derivations \citep{Frebel2013, Collet2018}.  In addition cool metal-poor giants are known to deviate from \ion{Fe}{1} and \ion{Fe}{2} ionization balance in 1D analyses, even when taking into account NLTE effects \citep{Karovicova2021}.  Therefore we try to avoid tuning these stellar parameters except in more extreme cases of disagreement from known trends.

Therefore, we first iterate on the atmospheric metallicity until it is in agreement with the average metallicity of good Fe lines. 
After converging on the atmospheric metallicity there is a good ionization balance of \ion{Fe}{1} and \ion{Fe}{2} for HD 122563 and GDR3 1760.  In addition the abundance trends of \ion{Fe}{1} line abundances with EP and EW are in reasonable agreement with expectations for 1D LTE analysis \citep{Collet2018}.  Given these results we do not alter the stellar parameters for these two stars any further (outside of updating the atmospheric \xm{C} and \xm{$\alpha$} with the abundance measurements).

For GDR3 0928, tuning the atmospheric metallicity alone does not appear to be sufficient.  Using the above determined parameters produces a large \logeps{\ion{Fe}{1}} $-$ \logeps{\ion{Fe}{2}} differences of $\sim 0.7$ dex, which is in the opposite sense of what might be expected in metal-poor stars \citep{Karovicova2021}, implying the stellar parameters need to be modified.  

While we have assumed enhancements in \xm{C} and \xm{$\alpha$}, the abundance of light elements in GDR3 0928 (including C, N, O, Na, Mg, and Si) exceeds even \xfe{X} $= +1$.  In this case elements like O and Mg will be the dominant electron donors and more important in driving the structure of the atmosphere in this star than Fe. So, for GDR3 0928 we tune the metallicity in the atmosphere to better match the \xh{X} of our light elements, primarily focusing on matching the O and Mg abundances.

The metallicity was adjusted until \xh{$\alpha$} matches average of the O and Mg abundances, while holding \xm{$\alpha$} $= +1$ which produces a better Fe ionization balance though only reducing the \logeps{\ion{Fe}{1}} $-$ \logeps{\ion{Fe}{2}} differences to $\sim 0.3-0.4$ dex.  Since these differences were still larger than the scatter in the \ion{Fe}{1} measurements and the trends with EP and EW exceeded expectations we modified the stellar parameters slightly ($\sim$ $-$70 K, $+$0.3 dex, $+$0.15 km/s) bringing the ionization balance and trends into better better agreement with those in HD 122563 and GDR3 1760.  

The resulting ionization balance is $\sim 0.15$ between the \ion{Fe}{1} and \ion{Fe}{2} for our final parameters for GDR3 0928, however, this difference is within the scatter seen in \ion{Fe}{1} lines.  Therefore, considering the small number of measurable \ion{Fe}{2} lines in GDR3 0928, the fact that cool giants may naturally deviate from ionization balance, and the significantly non-solar abundance ratios that are not consistently modeled in the atmospheres we use, we hesitate from fine tuning the parameters of this star further, as there may be many sources of systematic error that could account for these differences.

The final stellar parameters for each of these stars are reported in Table \ref{tab:params}.  For HD 122563 and GDR3 1760 we use the uncertainties from the initial stellar parameter calculations.  The uncertainties given for GDR3 0928 are estimated from the uncertainty in the slope of the \logeps{\ion{Fe}{1}} EP and EW trends, for the surface temperature and microturbulent velocity of GDR3 0928, respectively, and use the remaining ionization balance difference to estimate a \logg{}, uncertainty.  Finally, while the uncertainty in the atmospheric metallicity for GDR3 0928 is a poorly defined quantity, as with HD 122563 and GDR3 1760 we report the scatter in \ion{Fe}{1} abundances (rounded up) as a typical uncertainty, which is used to explore the systematic uncertainties in stellar parameters on the abundance measurements below. 

\subsection{Elemental Abundances}

\begin{deluxetable}{c r r r r r}
\tablewidth{0pt}
\tablecolumns{6}
\tablecaption{Line-by-line abundance measurements \label{tab:line_measurements}}
\tablehead{\colhead{Star} & \colhead{Species\tablenotemark{\scriptsize{a}}} & \colhead{Wavelength (\AA{})}& \colhead{EP\tablenotemark{\scriptsize{a}}} & \colhead{$\log gf$\tablenotemark{\scriptsize{b}}}& \colhead{$\log \epsilon$}}
\startdata
HD122563 & 106.0 &  4280.00 &       &        & 5.266 \\
HD122563 & 106.0 &  4304.00 &       &        & 5.282 \\
HD122563 & 106.0 &  4332.00 &       &        & 5.286 \\
GDR31760 & 106.0 &  4280.00 &       &        & 6.141 \\
GDR31760 & 106.0 &  4304.00 &       &        & 6.092 \\
GDR31760 & 106.0 &  4332.00 &       &        & 6.027
\enddata
\tablecomments{Table \ref{tab:line_measurements} is published in its entirety in the machine-readable format.  A portion is shown here for guidance regarding its form and content.}
\tablenotetext{a}{Atomic features are denoted in the format (Atmoic number).(ionization), with ionization $=$ 0 for neutral species, 1 for singly ionized species, etc. Molecular features are identified according to the atomic mass of each component (i.e., 106 for CH, 606 for C$_2$, and 607 for CN), and are not listed with an associated excitation potential.}
\tablenotetext{b}{Features for which there are multiple nearby lines are not specified with a $\log gf$ value (e.g., molecular features, lines with hyper-fine splitting, etc.)}
\end{deluxetable}

In Table \ref{tab:line_measurements} we report the lines used for our abundance measurements.  These are chosen as a combination of those used in \citet{Ji2016} for the sake of consistency in our comparison with other \rettwo{} stars, as well as lines used in other metal-poor star studies at redder wavelengths to take advantage of the wide wavelength range covered by GHOST \citep{Shetrone1996, Spite2018}.  We use minimum $\chi^2$ abundance measurements over equivalent width measurements because they are less susceptible to noise in the wings of weak lines, and therefore provide more precise abundance measurements.  Figure \ref{fig:cand_swanband} shows an example of our best fit to the 5172 \AA{} C$_2$ feature and Figure \ref{fig:cand_fits} shows some examples of the best fit syntheses for the lines of other elements.

Rejecting bad lines, produces the line-by-line abundance measurements given in Table \ref{tab:line_measurements}.  The final combined abundance measurements for each element and species are reported in Table \ref{tab:final_abundances}. Molecular and neutral species are given with abundances relative to \ion{Fe}{1} whereas ionized species are relative to \ion{Fe}{2}.

Comparing the ionization balance for elements with multiple species, and the abundances for different species and absorption systems of C and N can provide us another check on the quality of our stellar parameters and abundance measurements.  For ionization balance we have Fe and Ti, and also V for HD 122563.  We see a good ionization balance in Fe across all of our stars (with the above caveats for GDR3 0928).  For Ti and V we see differences on the order of 0.2-0.7 dex between the two species, however we note that below that particularly for \ion{Ti}{1} and \ion{V}{2} both are predominantly from low excitation potential lines where the NLTE corrections can be quite significant, several tenths of a dex.  

For GDR3 0928 we are able to measure C from both CH and C$_2$ features, and we find very good agreement between the two species especially considering the remaining uncertainty in its stellar atmosphere.  We do however, see relatively larger, 0.3 dex, differences between the N abundance we measure from the blue and red CN systems, that could suggest remaining errors in the stellar parameters of this star, or alternatively, could arise due to uncertainties in continuum placement, linelists, etc.

\rotate
\movetabledown=4in
\begin{deluxetable*}{c c c c c c c c c c c c c c c c c c c c c c}
\tablewidth{0pt}
\tablecolumns{22}
\tabletypesize{\scriptsize}
\tablecaption{Measured Abundances \label{tab:final_abundances}}
\tablehead{\colhead{} & \multicolumn{7}{c}{HD 122563} & \multicolumn{7}{c}{GDR3 1760} & \multicolumn{7}{c}{GDR3 0928} \\
\colhead{} & \colhead{} & \multicolumn{2}{c}{LTE} & \multicolumn{2}{c}{NLTE} & \colhead{} & \colhead{} & 
\colhead{} & \multicolumn{2}{c}{LTE} & \multicolumn{2}{c}{NLTE} & \colhead{} & \colhead{} & 
\colhead{} & \multicolumn{2}{c}{LTE} & \multicolumn{2}{c}{NLTE} & \colhead{} & \colhead{} \\ \cline{3-4} \cline{5-6} \cline{10-11} \cline{12-13} \cline{17-18} \cline{19-20}
\colhead{Species} & \colhead{N} & \colhead{$\log \epsilon$} & \colhead{\xfe{X}\tablenotemark{\scriptsize{a}}} & \colhead{$\log \epsilon$} & \colhead{\xfe{X}\tablenotemark{\scriptsize{a}}} & \colhead{$\sigma_{\rm disp}$} & \colhead{$\sigma_{\rm err}$} &
\colhead{N} & \colhead{$\log \epsilon$} & \colhead{\xfe{X}\tablenotemark{\scriptsize{a}}} & \colhead{$\log \epsilon$} & \colhead{\xfe{X}\tablenotemark{\scriptsize{a}}} & \colhead{$\sigma_{\rm disp}$} & \colhead{$\sigma_{\rm err}$} &
\colhead{N} & \colhead{$\log \epsilon$} & \colhead{\xfe{X}\tablenotemark{\scriptsize{a}}} & \colhead{$\log \epsilon$} & \colhead{\xfe{X}\tablenotemark{\scriptsize{a}}} & \colhead{$\sigma_{\rm disp}$} & \colhead{$\sigma_{\rm err}$}
}
\startdata
\ion{Fe}{1}  & 155 & 4.67 & -2.83 & 4.83 & -2.67 & 0.15 & 0.01 & 67 & 4.85 & -2.65 & 5.02 & -2.48 & 0.20 & 0.03 & 85 & 5 & -2.50 & 5.09 & -2.41 & 0.18 &  0.02 \\
\ion{Fe}{2}  & 13 & 4.73 & -2.77 & 4.73 & -2.77 & 0.07 & 0.04 & 5 & 4.85 & -2.65 & 4.86 & -2.64 & 0.26 & 0.09 & 3 & 4.85 & -2.65 & 4.86 & -2.64 & 0.07 &  0.10 \\
C (CH)  & 3 & 5.28 & -0.32 & 5.28 & -0.49 & 0.01 & 0.09 & 3 & 6.09 & 0.31 & 6.09 & 0.14 & 0.05 & 0.12 & 3 & 8.48 & 2.55 & 8.48 & 2.46 & 0.02 &  0.10 \\
C (C$_2$)  &  \nodata  &  \nodata  &  \nodata  &  \nodata  &  \nodata  &  \nodata  &  \nodata  &  \nodata  &  \nodata  &  \nodata  &  \nodata  &  \nodata  &  \nodata  &  \nodata  & 3 & 8.53 & 2.60 & 8.53 & 2.52 & 0.07 &  0.10 \\
N (CN$_{\rm B}$)  &  \nodata  &  \nodata  &  \nodata  &  \nodata  &  \nodata  &  \nodata  &  \nodata  &  \nodata  &  \nodata  &  \nodata  &  \nodata  &  \nodata  &  \nodata  &  \nodata  & 2 & 7.18 & 1.85 & 7.18 & 1.77 & 0.07 &  0.12 \\
N (CN$_{\rm R}$)  &  \nodata  &  \nodata  &  \nodata  &  \nodata  &  \nodata  &  \nodata  &  \nodata  &  \nodata  &  \nodata  &  \nodata  &  \nodata  &  \nodata  &  \nodata  &  \nodata  & 2 & 7.59 & 2.26 & 7.59 & 2.17 & 0.01 &  0.12 \\
\ion{O}{1}  & 1 & 6.77 & 0.91 & 6.77 & 0.75 & 0.00 & 0.15 &  \nodata  &  \nodata  &  \nodata  &  \nodata  &  \nodata  &  \nodata  &  \nodata  & 2 & 9.08 & 2.89 & 9.08 & 2.80 & 0.09 &  0.12 \\
\ion{Na}{1}  & 2 & 3.89 & 0.48 & 3.52 & -0.05 & 0.11 & 0.11 & 2 & 3.94 & 0.35 & 3.40 & -0.35 & 0.03 & 0.14 & 4 & 4.90 & 1.16 & 4.69 & 0.86 & 0.43 &  0.09 \\
\ion{Mg}{1}  & 9 & 5.37 & 0.6 & 5.39 & 0.46 & 0.08 & 0.03 & 6 & 5.31 & 0.37 & 5.34 & 0.22 & 0.14 & 0.06 & 7 & 6.70 & 1.60 & 6.68 & 1.49 & 0.35 &  0.13 \\
\ion{Si}{1}  & 7 & 5.23 & 0.55 & 5.23 & 0.39 & 0.19 & 0.07 &  \nodata  &  \nodata  &  \nodata  &  \nodata  &  \nodata  &  \nodata  &  \nodata  & 4 & 6.60 & 1.59 & 6.53 & 1.43 & 0.13 &  0.09 \\
\ion{K}{1}  & 1 & 2.83 & 0.63 & 2.48 & 0.12 & 0.00 & 0.15 & 2 & 3.15 & 0.77 & 2.85 & 0.3 & 0.14 & 0.14 & 2 & 3.44 & 0.91 & 3.09 & 0.47 & 0.08 &  0.12 \\
\ion{Ca}{1}  & 17 & 3.79 & 0.28 & 4.00 & 0.32 & 0.1 & 0.02 & 8 & 3.79 & 0.10 & 3.97 & 0.11 & 0.19 & 0.07 & 10 & 3.80 & -0.04 & 3.91 & -0.01 & 0.25 &  0.08 \\
\ion{Sc}{2}  & 10 & 0.44 & 0.07 & 0.44 & 0.06 & 0.08 & 0.03 & 2 & 0.75 & 0.25 & 0.75 & 0.24 & 0.25 & 0.14 &  \nodata  &  \nodata  &  \nodata  &  \nodata  &  \nodata  &  \nodata  &  \nodata \\
\ion{Ti}{1}  & 10 & 2.26 & 0.14 & 2.90 & 0.61 & 0.07 & 0.05 & 3 & 2.70 & 0.40 & 3.33 & 0.86 & 0.18 & 0.12 & 7 & 2.01 & -0.44 & 2.43 & -0.11 & 0.26 &  0.14 \\
\ion{Ti}{2}  & 35 & 2.47 & 0.29 & 2.56 & 0.38 & 0.13 & 0.02 & 13 & 2.94 & 0.64 & 2.96 & 0.65 & 0.37 & 0.10 & 10 & 2.59 & 0.29 & 2.58 & 0.26 & 0.19 &  0.12 \\
\ion{V}{1}  & 1 & 1.01 & -0.09 & 1.01 & -0.25 & 0.00 & 0.15 &  \nodata  &  \nodata  &  \nodata  &  \nodata  &  \nodata  &  \nodata  &  \nodata  & 1 & 0.80 & -0.63 & 0.80 & -0.72 & 0.00 &  0.18 \\
\ion{V}{2}  & 1 & 1.35 & 0.19 & 1.35 & 0.19 & 0.00 & 0.15 &  \nodata  &  \nodata  &  \nodata  &  \nodata  &  \nodata  &  \nodata  &  \nodata  &  \nodata  &  \nodata  &  \nodata  &  \nodata  &  \nodata  &  \nodata  &  \nodata \\
\ion{Cr}{1}  & 12 & 2.46 & -0.35 & 2.96 & -0.01 & 0.14 & 0.04 & 2 & 2.58 & -0.40 & 2.99 & -0.17 & 0.16 & 0.14 & 6 & 2.65 & -0.49 & 3.04 & -0.19 & 0.20 &  0.08 \\
\ion{Mn}{1}  & 6 & 2.05 & -0.55 & 2.29 & -0.47 & 0.16 & 0.07 &  \nodata  &  \nodata  &  \nodata  &  \nodata  &  \nodata  &  \nodata  &  \nodata  & 1 & 2.40 & -0.53 & 2.61 & -0.41 & 0.00 &  0.18 \\
\ion{Co}{1}  & 4 & 2.30 & 0.14 & 3.07 & 0.75 & 0.12 & 0.08 &  \nodata  &  \nodata  &  \nodata  &  \nodata  &  \nodata  &  \nodata  &  \nodata  &  \nodata  &  \nodata  &  \nodata  &  \nodata  &  \nodata  &  \nodata  &  \nodata \\
\ion{Ni}{1}  & 6 & 3.63 & 0.24 & 3.63 & 0.07 & 0.10 & 0.04 & 1 & 3.63 & 0.06 & 3.63 & -0.11 & 0.00 & 0.20 & 3 & 3.48 & -0.24 & 3.48 & -0.33 & 0.12 &  0.10 \\
\ion{Zn}{1}  & 2 & 2.00 & 0.27 & 2.00 & 0.10 & 0.02 & 0.11 &  \nodata  &  \nodata  &  \nodata  &  \nodata  &  \nodata  &  \nodata  &  \nodata  &  \nodata  &  \nodata  &  \nodata  &  \nodata  &  \nodata  &  \nodata  &  \nodata \\
\ion{Sr}{2}  & 2 & 0.11 & 0.01 & 0.11 & 0.00 & 0.36 & 0.11 & 1 & 0.99 & 0.78 & 0.99 & 0.76 & 0.00 & 0.2 &  \nodata  &  \nodata  &  \nodata  &  \nodata  &  \nodata  &  \nodata  &  \nodata \\
\ion{Y}{2}  & 5 & -0.83 & -0.27 & -0.83 & -0.27 & 0.07 & 0.07 & 1 & 0.07 & 0.52 & 0.07 & 0.5 & 0.00 & 0.20 & 3 & 0.62 & 1.06 & 0.62 & 1.05 & 0.06 &  0.10 \\
\ion{Zr}{2}  & 4 & -0.12 & 0.07 & -0.12 & 0.07 & 0.10 & 0.08 &  \nodata  &  \nodata  &  \nodata  &  \nodata  &  \nodata  &  \nodata  &  \nodata  & 1 & 1.59 & 1.66 & 1.59 & 1.65 & 0.00 &  0.18 \\
\ion{Ba}{2}  & 5 & -1.51 & -0.92 & -1.42 & -0.83 & 0.05 & 0.07 & 3 & 0.76 & 1.24 & 0.57 & 1.03 & 0.17 & 0.12 & 3 & 1.11 & 1.57 & 0.94 & 1.39 & 0.13 &  0.10 \\
\ion{La}{2}  & 3 & -2.59 & -0.92 & -2.59 & -0.93 & 0.05 & 0.09 &  \nodata  &  \nodata  &  \nodata  &  \nodata  &  \nodata  &  \nodata  &  \nodata  & 1 & 0.91 & 2.45 & 0.91 & 2.44 & 0.00 &  0.18 \\
\ion{Ce}{2}  & 2 & -1.99 & -0.80 & -1.99 & -0.80 & 0.16 & 0.11 & 1 & 0.57 & 1.64 & 0.57 & 1.63 & 0.00 & 0.20 & 1 & 0.85 & 1.92 & 0.85 & 1.91 & 0.00 &  0.18 \\
\ion{Nd}{2}  & 6 & -1.93 & -0.57 & -1.93 & -0.58 & 0.23 & 0.09 &  \nodata  &  \nodata  &  \nodata  &  \nodata  &  \nodata  &  \nodata  &  \nodata  & 2 & 1.14 & 2.37 & 1.14 & 2.36 & 0.29 &  0.12 \\
\ion{Sm}{2}  &  \nodata  &  \nodata  &  \nodata  &  \nodata  &  \nodata  &  \nodata  &  \nodata  & 1 & -0.02 & 1.68 & -0.02 & 1.66 & 0.00 & 0.20 & 5 & 0.68 & 2.37 & 0.68 & 2.36 & 0.11 &  0.08 \\
\ion{Eu}{2}  & 2 & -2.78 & -0.53 & -2.78 & -0.53 & 0.05 & 0.11 & 5 & -0.29 & 1.85 & -0.29 & 1.83 & 0.14 & 0.09 & 3 & 0.24 & 2.36 & 0.24 & 2.35 & 0.14 &  0.10 \\
\ion{Dy}{2}  &  \nodata  &  \nodata  &  \nodata  &  \nodata  &  \nodata  &  \nodata  &  \nodata  &  \nodata  &  \nodata  &  \nodata  &  \nodata  &  \nodata  &  \nodata  &  \nodata  & 1 & 1.22 & 2.77 & 1.22 & 2.76 & 0.00 &  0.18
\enddata
\tablenotetext{a}{Reporting \feh{} instead of \xfe{X} for \ion{Fe}{1} and \ion{Fe}{2}}
\end{deluxetable*}

\subsection{Abundance Uncertainties}

In Table \ref{tab:final_abundances} we also report the standard deviation, $\sigma_{\rm stdev}$, of all of the lines of each species, and our estimated statistical uncertainty, for which we report the standard error of the mean from all of the lines we have measured, $\sigma_{\rm err} = \sigma_{\rm stdev}/\sqrt{\rm N_{\rm lines}}$.  For species with fewer than five lines, instead of using scatter from the lines of that species, we use the scatter from the \ion{Fe}{1} lines as a better estimate of typical line-to-line scatter.  The standard deviations and standard errors of the mean only estimate random errors on line measurements, and other systematics, such as systematic errors in stellar parameters or in line choice (for species with few lines), may not be well captured by these numbers.

 \startlongtable
\begin{deluxetable}{c r r r r}
\tablewidth{0pt}
\tablecolumns{5}
\tablecaption{Systematic Errors \label{tab:sys_errors}}
\tablehead{\colhead{Species} & \colhead{$+\Delta_{\rm Teff}$} & \colhead{$+\Delta_{\rm logg}$} & \colhead{$+\Delta_{\rm [M/H]}$} & \colhead{$+\Delta_{\rm micro}$}}
\startdata
\hline
\multicolumn{5}{c}{HD 122563} \\
\hline
CH & 0.19 & -0.01 & -0.02 & -0.01 \\
\ion{O}{1} & 0.08 & 0.04 & 0.00 & 0.03 \\
\ion{Na}{1} & 0.18 & -0.00 & -0.05 & -0.05 \\
\ion{Mg}{1} & 0.09 & -0.01 & -0.01 & -0.02 \\
\ion{Si}{1} & 0.05 & 0.00 & -0.00 & -0.00 \\
\ion{K}{1} & 0.09 & -0.00 & -0.01 & -0.01 \\
\ion{Ca}{1} & 0.08 & -0.00 & -0.01 & -0.01 \\
\ion{Sc}{2} & 0.06 & 0.02 & -0.02 & 0.01 \\
\ion{Ti}{1} & 0.14 & -0.00 & -0.01 & -0.01 \\
\ion{Ti}{2} & 0.04 & 0.02 & -0.02 & 0.01 \\
\ion{V}{1} & 0.11 & -0.00 & 0.00 & -0.01 \\
\ion{V}{2} & 0.03 & 0.03 & -0.00 & 0.01 \\
\ion{Cr}{1} & 0.14 & -0.00 & -0.02 & -0.02 \\
\ion{Mn}{1} & 0.15 & -0.00 & -0.02 & -0.03 \\
\ion{Fe}{1} & 0.13 & -0.00 & -0.02 & -0.02 \\
\ion{Fe}{2} & 0.00 & 0.02 & -0.02 & 0.01 \\
\ion{Co}{1} & 0.15 & -0.00 & -0.01 & -0.01 \\
\ion{Ni}{1} & 0.10 & -0.00 & -0.00 & -0.01 \\
\ion{Zn}{1} & 0.04 & 0.01 & -0.00 & 0.01 \\
\ion{Sr}{2} & 0.08 & 0.02 & -0.05 & 0.00 \\
\ion{Y}{2} & 0.06 & 0.02 & -0.00 & 0.01 \\
\ion{Zr}{2} & 0.06 & 0.02 & -0.00 & 0.01 \\
\ion{Ba}{2} & 0.07 & 0.02 & -0.01 & 0.01 \\
\ion{La}{2} & 0.07 & 0.01 & 0.00 & 0.04 \\
\ion{Ce}{2} & 0.03 & 0.01 & 0.00 & 0.02 \\
\ion{Nd}{2} & 0.09 & 0.02 & 0.00 & 0.01 \\
\ion{Eu}{2} & 0.08 & 0.03 & 0.00 & 0.02 \\
\ion{Dy}{2} & 0.16 & 0.05 & -0.02 & -0.37 \\
\hline
\multicolumn{5}{c}{GDR3 1760} \\
\hline
CH & 0.25 & -0.05 & -0.04 & 0.02 \\
\ion{Na}{1} & 0.13 & -0.01 & -0.07 & -0.00 \\
\ion{Mg}{1} & 0.11 & -0.02 & -0.02 & 0.00 \\
\ion{K}{1} & 0.09 & -0.01 & -0.02 & 0.00 \\
\ion{Ca}{1} & 0.11 & -0.01 & -0.02 & 0.01 \\
\ion{Sc}{2} & 0.09 & 0.03 & -0.03 & 0.01 \\
\ion{Ti}{1} & 0.13 & -0.01 & -0.02 & 0.00 \\
\ion{Ti}{2} & 0.06 & 0.04 & -0.03 & 0.01 \\
\ion{Cr}{1} & 0.15 & -0.01 & -0.03 & 0.01 \\
\ion{Fe}{1} & 0.13 & -0.00 & -0.03 & 0.00 \\
\ion{Fe}{2} & 0.02 & 0.05 & -0.03 & 0.00 \\
\ion{Ni}{1} & 0.12 & -0.00 & -0.03 & 0.00 \\
\ion{Sr}{2} & 0.10 & -0.01 & -0.04 & -0.00 \\
\ion{Y}{2} & 0.07 & 0.04 & -0.01 & 0.00 \\
\ion{Ba}{2} & 0.11 & 0.03 & -0.09 & -0.00 \\
\ion{Ce}{2} & 0.08 & 0.04 & -0.02 & 0.01 \\
\ion{Sm}{2} & 0.06 & 0.04 & -0.02 & 0.03 \\
\ion{Eu}{2} & 0.09 & 0.05 & -0.01 & 0.01 \\
\hline
\multicolumn{5}{c}{GDR3 0928} \\
\hline
CH & -0.06 & 0.04 & 0.02 & 0.02 \\
C$_2$ & -0.00 & 0.02 & 0.01 & 0.01 \\
CN Blue & -0.02 & 0.02 & 0.03 & 0.04 \\
CN Red & -0.08 & 0.10 & 0.05 & 0.09 \\
\ion{O}{1} & 0.01 & 0.08 & 0.05 & 0.07 \\
\ion{Na}{1} & 0.12 & -0.10 & -0.06 & -0.11 \\
\ion{Mg}{1} & 0.13 & -0.08 & -0.04 & -0.09 \\
\ion{Si}{1} & -0.05 & 0.06 & 0.04 & 0.05 \\
\ion{K}{1} & 0.12 & 0.01 & 0.01 & -0.03 \\
\ion{Ca}{1} & 0.09 & 0.01 & 0.00 & 0.00 \\
\ion{Ti}{1} & 0.14 & 0.03 & 0.02 & 0.01 \\
\ion{Ti}{2} & -0.00 & 0.09 & 0.05 & 0.07 \\
\ion{V}{1} & 0.14 & 0.04 & 0.02 & 0.01 \\
\ion{Cr}{1} & 0.13 & 0.04 & 0.03 & 0.01 \\
\ion{Mn}{1} & 0.08 & -0.00 & 0.02 & 0.03 \\
\ion{Fe}{1} & 0.07 & 0.03 & 0.02 & 0.02 \\
\ion{Fe}{2} & -0.07 & 0.13 & 0.09 & 0.11 \\
\ion{Ni}{1} & 0.05 & 0.04 & 0.02 & 0.04 \\
\ion{Y}{2} & -0.00 & 0.08 & 0.04 & 0.07 \\
\ion{Zr}{2} & 0.05 & 0.10 & 0.01 & 0.09 \\
\ion{Ba}{2} & 0.02 & 0.07 & 0.04 & 0.04 \\
\ion{La}{2} & -0.03 & 0.12 & 0.06 & 0.11 \\
\ion{Ce}{2} & -0.01 & 0.08 & 0.04 & 0.07 \\
\ion{Nd}{2} & 0.03 & 0.13 & 0.08 & 0.11 \\
\ion{Sm}{2} & 0.01 & 0.06 & 0.04 & 0.04 \\
\ion{Eu}{2} & 0.01 & 0.10 & 0.05 & 0.08 \\
\ion{Dy}{2} & -0.11 & 0.07 & 0.04 & 0.09 \\
\enddata
\end{deluxetable}

To investigate some systematic sources of error/uncertainty, we  recalculate our abundances but by changing each of our stellar parameters by +1$\sigma$ using the uncertainties given in Table \ref{tab:params}.  The differences between these abundances and our final reported abundances (along with their sign) are given in Table \ref{tab:sys_errors}.

While the systematic uncertainties are relatively larger for GDR3 0928, they are small enough that the significant enhancement in light elements and neutron capture elements should be robust to stellar parameter uncertainties.  While we should be very careful about drawing detailed conclusions from the abundances in GDR3 0928, it does appear that its unusual abundance ratios are real and that this is an interesting star to investigate more closely.
 
\subsection{NLTE Effects}

Non-Local Thermodynamic Equilibrium (NLTE) effects can be considerable in metal-poor stars.  We use the online INSPECT\footnote{\url{http://inspect-stars.com}} tool to calculate Na and Mg NLTE corrections \citep{Lind2011na, Osorio2015mg, Osorio2016mg}, the online MPIA NLTE\footnote{\url{https://nlte.mpia.de/gui-siuAC_secE.php}} tool for Si, Ca, Cr, Ti, Mn, and Fe \citep{Mashonkina2007ca, Bergemann2010cr, Bergemann2011ti, Bergemann2012fe, Bergemann2013si, Bergemann2019mn}, the Mashonkina online tables\footnote{\url{http://www.inasan.ru/~lima/pristine/}} for Sr and Ba NLTE corrections \citep{Belyakova1997, Mashonkina2019}, and \citet{Ivanova2000} for K.

In INSPECT we use the stellar parameters given in Table \ref{tab:params}, except for an $\alpha$-scaled metallicity value for GDR3 0928, \xh{$\alpha$} $=-0.3$, along with the measured \logeps{X} in Table \ref{tab:line_measurements}.  Mg NLTE corrections are available from INSPECT and MPIA \citep[][]{Bergemann2017mg} but we opt for the INSPECT corrections because it allows for non-solar Mg abundance ratios. Fe NLTE corrections are also offered by INSPECT and MPIA and provide very similar results, so we use the MPIA corrections, because they are available for more lines.

For the MPIA corrections, we use the spherical 1D model option (except for Ca and Mn, for which the spherical models are not available so we use the 1D plane-parallel atmospheres).  The MPIA NTLE corrections assume solar-scaled abundance ratios, so the metallicity was altered for GDR3 0928 depending on the element.  

Lighter elements like Mg and Si, have abundances consistent with \xh{$\alpha$} that dominates the relevant atmospheric metallicity for this star, so NLTE calculations for these elements were done with \xh{$\alpha$} $=-0.3$.  The heavier elements are more discrepant since their abundance does not match the atmospheric metallicity, giving us two options for the NLTE corrections:  1) using the abundance of the heavier elements as the input metallicity (so the NLTE corrections match the correct line abundance) or 2) use our best estimate of the metallicity to better match the the correct atmospheric structure.

We compare the NLTE corrections for these two options by considering their impact on the ionization balance for Ti and Fe, and find that using a metallicity that better matches the abundance of these heavier elements \xh{M} $= -2.5$ produces a slightly better ionization balance.  Nonetheless \ion{Ti}{1} - \ion{Ti}{2} is still negative for GDR3 0928, whereas it is positive HD 122563 and GDR3 1760.  It is unclear if this inconsistency lies in errors in the stellar parameters or if the NLTE corrections are still not large enough in this star, because that the \ion{Ti}{1} lines have a low EP and large NLTE corrections.

Estimates of the Sr and Ba NLTE corrections were obtained from the Mashonkina online tables, using the \xfe{Sr} and \xfe{Ba} $= +0.5$ tables for GDR3 0928 and GDR3 1760, and the \xfe{Sr} $= 0.0$ and \xfe{Ba} $= -0.5$ for HD 122563.  We elect the nlte correction for the nearest point in \teff, \logg, and \xh{M}, for each star.  Again for GDR3 0928 we use the highest metallicity point available, \xh{M} $= -2.0$ to better match its Ba abundance. Finally for K we estimate rough NLTE corrections from \citet{Ivanova2000}, referencing the \xh{M} $= -2.0$ giant model at the temperatures of our stars \citep[but note that the true NLTE corrections will depend on the abundance of K;][]{Andrievsky2010}.

\begin{deluxetable}{c r r r r}
\tablewidth{0pt}
\tablecolumns{12}
\tablecaption{NLTE Corrections \label{tab:nlte_corrections}}
\tablehead{\colhead{Star} & \colhead{Species\tablenotemark{\scriptsize{a}}} & \colhead{Wavelength (\AA{})}& \colhead{NLTE Correction} & \colhead{Source\tablenotemark{\scriptsize{b}}}}
\startdata
HD122563 & 11.0 & 5889.95 & -0.355 & 1 \\
HD122563 & 11.0 & 5895.92 & -0.385 & 1 \\
GDR31760 & 11.0 & 5889.95 & -0.563 & 1 \\
GDR31760 & 11.0 & 5895.92 & -0.502 & 1 \\
GDR30928 & 11.0 & 5889.95 & 0.006 & 1 \\
GDR30928 & 11.0 & 5895.92 & -0.153 & 1
\enddata
\tablecomments{Table \ref{tab:nlte_corrections} is published in its entirety in the machine-readable format.  A portion is shown here for guidance regarding its form and content.}
\tablenotetext{a}{The species are denoted in the format (Atmoic number).(ionization), with ionization $=$ 0 for neutral species, 1 for singly ionized species, etc. }
\tablenotetext{b}{1 -- INSPECT; 2 -- MPIA; 3 -- \citep{Ivanova2000}; 4 -- Mashonkina online table}
\end{deluxetable}

The NLTE corrections we use are given in Table \ref{tab:nlte_corrections}.  Since they are not available for all of the lines that we measure, we average the NLTE corrections for each species and apply that average correction to our line-averaged abundances to obtain the NLTE abundances shown in Table \ref{tab:final_abundances}.  Not all elements have NLTE corrections but we repeat those that do not because of the Fe NLTE corrections that factor into their \xfe{X} ratios.

\begin{figure*}
\centering\includegraphics[scale=0.43,trim = 0.in 0.in 0.in 0.in, clip]{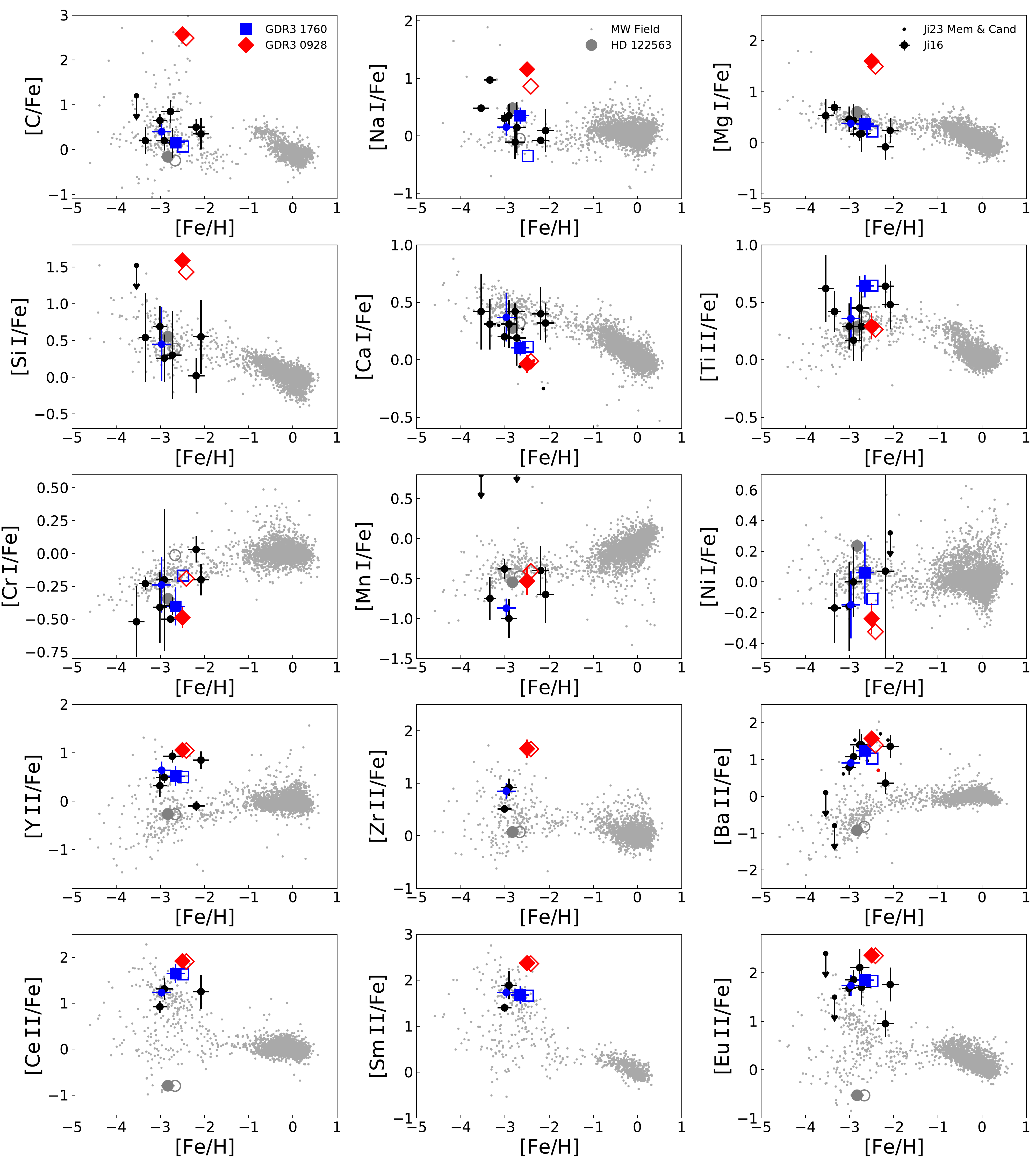}
  \caption{LTE \xfe{X} abundances for a selection of elements as a function of metallicity. Symbols are for HD 122563 (gray circle), GDR3 1760 (blue square), and GDR3 0928 (red diamond), compared to other Reticulum 2 stars (black points) from \citet[large points][]{Ji2016} and \citet[small points][]{Ji2016}, preferring the abundances from \citet{Ji2016} for stars in common, except for GDR3 1760 and GDR3 0928, whose measurements are all shown when possible.  Our NLTE abundances for HD 122563, GDR3 1760, and GDR3 0928 are also shown (open symbols).}
  \label{fig:xfe}
\end{figure*}

\section{Results}
\label{sec:results}

\begin{figure*}[t]
  \centering
  \includegraphics[scale=0.4,trim = 0.in 0.in 0.in 0.in, clip]{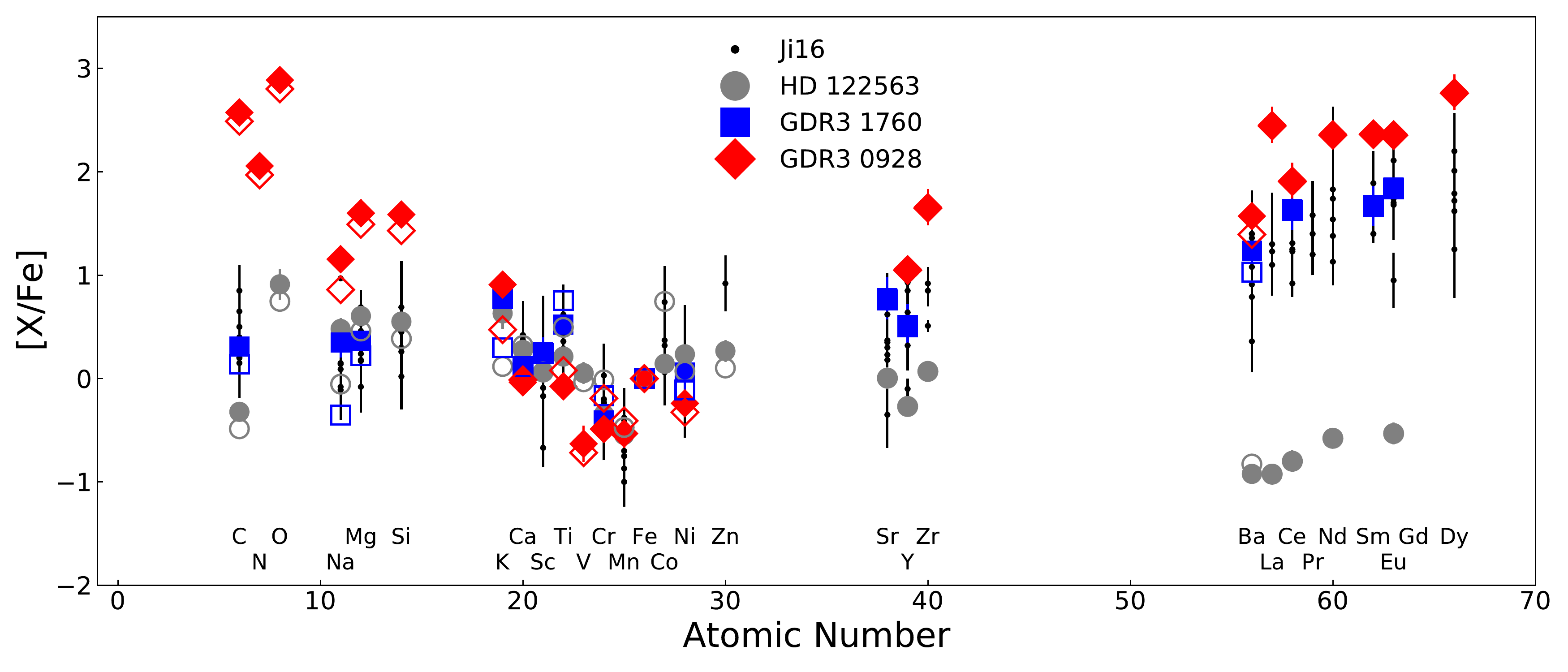}
  \caption{\xfe{X} abundances in LTE (filled points) and with NLTE corrections applied (open points) as a function of atomic number for HD 122563, GDR3 1760, and GDR3 0928.  As seen in other \rettwo{} stars GDR3 1760 and GDR3 0928 both exhibit neutron capture element abundances that are enhanced over field stars with a typical r-process dominated abundance pattern.  GDR3 0928 also has enhanced light element abundances that are not seen in other \rettwo{} stars to date.}
  \label{fig:abund_patterns}
\end{figure*}

The chemical abundances from our analysis of the two \rettwo{} stars, GDR3 0928 and GDR3 1760, and our benchmark star HD 122563, are compared in Figure \ref{fig:xfe}. 
We also compare to other \rettwo{} stars analyzed by \citet{Ji2016, Ji2023}, including both their ``members'' and ``candidates'', and a sample of Milky Way field stars compiled from the literature \citep{Chen2000,Mashonkina2004,Reddy2006,Adibekyan2012, Mishenina2013,Bensby2014,Roederer2014,Battistini2015, Battistini2016,Brewer2016,daSilva2015,Delgadomena2017, Jonsson2017,Duong2019,Forsberg2019,Lomaeva2019}.
Both the LTE and NLTE abundance results are both provided in Figure \ref{fig:xfe}, because many of the results in the literature do not include NLTE corrections. The abundances for the stars in our sample are also shown as \xfe{X} ratios as a function of atomic number
in Figure \ref{fig:abund_patterns}. 

Examination of Figure \ref{fig:xfe} shows that our results for the benchmark HD 122563 are in good agreement with the Milky Way field stars, and GDR3 1760 has similar chemical abundances to the other known \rettwo{} stars.  In particular, GDR3 1760 displays the enhancements in neutron-capture elements previously found by \citet{Ji2016}.  Our redder target GDR3 0928 has Fe-peak abundances and neutron-capture enhancements similar to other stars in \rettwo{}, but it uniquely shows strong enhancements in the light elements (C, N, O, Na, Mg, and Si).  Those abundance ratios appear to decrease with increasing atomic number, such that [C/Fe] is more enhanced than [Si/Fe].  The N and Na abundances may imply a slight odd-even effect, which may indicate contributions from one or only a few massive supernovae (discussed further below).

\subsection{Light elements (C, N)}

Previous studies have shown that GDR3 1760 is slightly enhanced in C, making it one of the few stars in \rettwo{} to be CEMP stars \citep{Ji2016, Roederer2016}.  
In GDR3 0928, we measure C and N from the CH g band, C$_2$ Swan bands, and the red and blue CN features, which all suggest high abundances of C and N in this star.  The very high \xfe{C} $= +2.6$ identifies GDR3 0928 as a CEMP star, by any definition in the literature \citep[e.g., \xfe{C} $> 1.0$ or $> 0.7$ from][respectively]{Beers2005,Aoki2007}.  

Internal mixing alters the surface abundances of C and N as stars evolve up the red giant branch, where a deepening surface convection zone can reach the H-burning layers affected by CNO-cycling and dredge up this processed material to the surface.  The impact on C is small, \xfe{C} $\approx -0.05$, but N can increase by $>0.5$ dex (also see mixing calculator in \citealt{Placco2014}).  The initial values of N are not constrained, however in GDR3 0928, our values of \xfe{C} and \xfe{O} are so high that they may reflect contributions from faint SNe  (discussed below).  Another possibility could be the third dredge up that occurs during Asymptotic Giant Branch evolution, if the surface convection zone has reached He-burned layers (raising C and O from the triple-$\alpha$ process to the surface, e.g., \citealt{Herwig2005}). The CMD position of this star in Figure \ref{fig:targeting} argues against this possibility though, as it appears closer to the RGB bump (on its first ascent of the giant branch) than the upper AGB (third dredge-up). 
 
\subsection{$\alpha$-elements}

As is typical of halo and dwarf UFD stars, HD 122563 and GDR3 1760 show slightly to moderately super-solar \xfe{$\alpha$} in the $\alpha$-elements measured here (O, Mg, Si, Ca, and Ti).  
Alternatively, GDR3 0928 features similar abundances to other \rettwo{} stars in the heavier, ``explosive'' $\alpha$-elements (Ca and Ti) consistent with those found by \citep{Ji2023}.  However, for the lighter, ``hydrostatic'' $\alpha$-elements (O, Mg, and Si), we find that GDR3 0928 is significantly enhanced with \xfe{X} $\gtrsim +1$.
 
We note that our Mg abundances for this star, GDR3 0928, show a relatively large dispersion.  This is driven by the Mg b lines, which provide much lower abundances than from the other Mg lines (by $\gtrsim$ 0.5 dex).  The Mg b lines are collisionally broadened and extremely sensitive to the stellar parameters, including surface gravity and microturbulence.  If we exclude these lines, our Mg abundance for GDR3 0928 is raised to \xfe{Mg} $=$ +1.81 (LTE).

\subsection{Odd-Z Elements (Na, K)}

Again, our analysis shows that HD 122563 and GDR3 1760 show slightly to moderately super-solar \xfe{odd-Z} ratios in the odd-Z elements measured here (Na and K).  
Alternatively,
GDR3 0928 is enhanced in Na and K, particularly in comparison to our analysis of GDR3 1760 and the other \rettwo{} stars in the literature.  For Na, this remains true even after correcting for NLTE effects.
There is a hint of an odd-even effect in that [Na/Fe] (A=11) has a lower abundance than [Mg/Fe] (A=12) and [Si/Fe] (A=14), however this does not extend to [K/Fe] (A=19) which is higher than [Ca/Fe] (A=20).

\subsection{Fe-peak elements}
The measurements of the Fe-peak elements in our stars are fairly unremarkable, i.e., typical of Milky Way halo and UFD stars.  This suggests standard production of Fe-peak elements in CCSN sampled over a typical range in the IMF.

\subsection{Neutron capture elements}
\label{sec:ncap}
 
Both \rettwo{} stars GDR3 0928 and GDR3 1760 are strongly enhanced in neutron capture elements, like the other \rettwo{} stars in the literature.  Based on \xfe{Eu} $= +2.4$ and [Ba/Eu] $= -0.8$, we identify GDR3 0928 as an r-II star \citep{Christlieb2004}, and, given its carbon enhancement, we identify this as a CEMP-r star. 

\begin{figure}
  \centering
  \includegraphics[scale=0.3,trim = 0.in 0.in 0.in 0.in, clip]{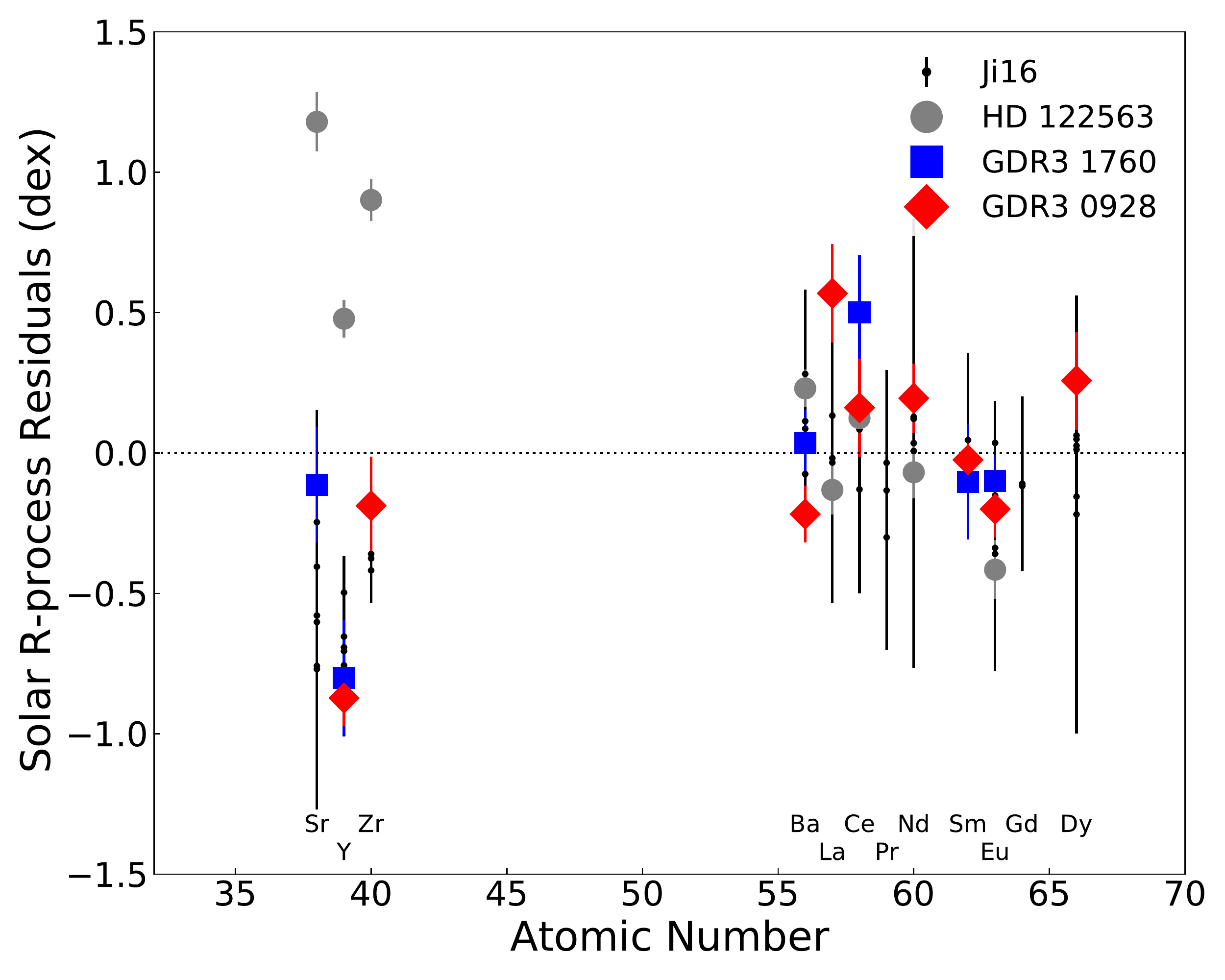}
  \caption{Residual abundances (using LTE) from a scaled solar r-process abundance pattern in HD 122563, GDR3 1760 and GDR3 0928, fit to the heavier neutron capture elements (see text for details).}
  \label{fig:r_proc_resid}
\end{figure}

We examine the r-process {\it pattern} by
following the formulation given in \citet{Ji2016} comparing the neutron capture ratios to the solar r-process pattern \citep[adopting solar values from][]{Prantzos2020}: 
\begin{equation}
    \log \epsilon(\mathrm{X}_{*}) = \log \epsilon(\mathrm{X}_{\odot, \mathrm{r-proc}}) + \epsilon_{\mathrm{offset}}
\end{equation}
Here, $\epsilon_{\mathrm{offset}}$ is the average \xh{r} for each star. The residuals from this scaled solar r-process are shown in Figure \ref{fig:r_proc_resid}, which also includes the residuals of the heavy element abundances measured in \rettwo{} stars by \citet{Ji2016}.  The r-process residuals for our two \rettwo{} stars are sufficiently close to zero to imply they are purely enriched by the r-process.

\citet{Ji2016} also compare their r-process patterns with the metal-poor r-process rich benchmark star, CS22892-052 \citep{Sneden2003}. They found their offsets in Y and Zr (A = 39 and 40) are likely due to slight differences in the r-process abundance pattern of the Sun compared to that of r-II metal-poor stars.  We also find low abundances of Y in our two \rettwo{} stars, also assumed to be due to the impure r-process solar abundance of this element.

Therefore, GDR3 0928 has an r-process pattern consistent with the other r-process enhanced \rettwo{} stars. 
It may appear to be more enhanced in an absolute sense 
(as seen from its Zr, Ce, Sm and Eu abundances in Figure \ref{fig:xfe}),  
however this could be due to an incomplete mixing of the ejecta from the r-process event (e.g., see \citealt{McWilliam1997, Burris2000}).  We note that these abundances are also within one sigma errors of the same enrichment given the impact of the stellar parameter uncertainties on the abundances.

\subsection{Comparison with Literature}

While a full comparison of HD 122563 with the literature is beyond the scope of this work, we note that the abundances we find are in good agreement with typical values measured for HD 122563, even the low neutron capture ratio \citep[e.g.,][]{Honda2006, Collet2018, Kielty2021}.  Additionally when accounting for uncertainties and systematic differences due to the choice of stellar parameters used, we also find a reasonable agreement between this work and \citet{Ji2016} for GDR3 1760, both of which are shown in Figure \ref{fig:xfe}. 

For GDR3 0928, \citet{Ji2023} provide a limited number of abundances, i.e., \xh{Fe} $= -2.35$, \xh{Ca} $= -2.36$, and \xh{Ba} $= -1.64$ (from $\lambda$6496 \AA{} only), all in LTE.  Our Fe and Ca measurements of \xh{Fe} $= -2.5$, \xh{Ca} $= -2.54$ are relatively similar, however, our \xh{Ba} $= -0.93$ is significantly higher. Accounting for isotopic and hyperfine structure components, only has a maximum impact of $\Delta$(Ba)$<+0.1$ dex for our Ba II lines ($\lambda\lambda$5853, 6141, 6496 \AA).  
The difference in Ba is likely due to the differences in stellar parameters. They used stellar parameters very close to our initial parameters, for which we found relatively large $\sim 0.7$ dex differences between neutral and ionized species of Fe. By adjusting these to achieve a better ionization balance, then considerable change was found in the abundances of other ionized species, like \ion{Ba}{2}. These changes can account for most of the difference seen between these two measurements.  

\section{Discussion}
\label{sec:discussion}

As \rettwo{} is an UFD that appears to have undergone relatively little star formation, it is a unique system that can also be used to study metal mixing and enrichment in a very low mass system at very early epochs.  Detailed spectral analyses of old stars in classical dwarf galaxies and UFDs has revealed that there is inhomogeneous mixing in these systems \citep[e.g.,][]{Venn2012, Leaman2012, Norris2017}, which is also supported by in low mass galaxy simulations \citep[e.g.,][]{Revaz2012, Romano2015, Corlies2018, Applebaum2020, Applebaum2021}.
Recent analyses of the two UFDs, Car II and Ret II, by \citet{Alexander2023} have incorporated inhomogeneous mixing as driven by SN explosions.  Chemical enrichment from the supernova bubbles spreads in the interstellar medium, causing dispersion in the elemental abundances, and driving galactic outflows that quench the star formation activity at early times. For Ret II, they predict a fairly large average outflow mass-loading factor, $\sim10^3$.

Simultaneously, \citep{Ji2023} have found that r-process-enhanced stars in \rettwo{} have a mean [Ba/H]= $-1.68 \pm 0.07$ and unresolved intrinsic dispersion $\sigma$[Ba/H]$<0.20$. This implies that \rettwo{}'s r-process rich material was well mixed before the formation of the r-process enhanced stars (also see \citealt{Brauer2019}).  They suggest mixing would require at least 100 Myr before the later r-process enrichment occurred, a sufficiently short time that a prompt event is required, such as a collapsar disk wind or prompt neutron star merger.

\subsection {New star in \rettwo{}: GDR3 0928}

We have presented new spectral observations for the bright, red star GDR3 0928.  This star has a similar radial velocity and low metallicity ([Fe/H]=$-2.5$) to the other stars in \rettwo{}, suggesting that it is a bona fide member of this UFD.  We have found that it is enriched in r-process elements like the majority of the giants (72\%, \citealt{Ji2023}) in \rettwo{}, at such levels that identify it as an r-II star.  In addition, it is enriched in carbon, identifying it is a CEMP-r star.

Unlike most other r-process rich stars in \rettwo{}, GDR3 0928 is also strongly enhanced in several light elements: N, O, Na, Mg, and Si, in addition to C.  To interpret this abundance pattern, we consider the other stars analysed in \rettwo{}.

First, \citet{Roederer2016} and \citet{Ji2016} also identified a C-enhanced star, \xfe{C}$_{\rm corrected} > 1$ that shows an r-process enrichment, at [Fe/H]$=-3$ (DES J033523-540407). However, there is a large range of C abundances among the r-process rich stars in \rettwo{}.  The lack of correlation between between these two enrichments (C and r) suggests that the processes producing them are unrelated. 

Second, two stars are identified with moderate enhancements in the light $\alpha$-elements
(Na, Mg, and possibly Si), both at [Fe/H]$<-3.3$ (DES J033531-540148 and DES J033556-540316).  These two stars show no r-process enrichments, and either have low C abundances or upper limits.  Furthermore, none of the higher metallicity r-process rich stars show the C nor light $\alpha$-element enrichments (of the magnitude seen in GDR3 0928), which suggests that the light $\alpha$-element enhancements are also decoupled from the r-process enrichment sources. The lack of other stars with these enrichments suggests the source of the light $\alpha$-elements in GDR3 0928 is not a dominant contributor to the chemical evolution of \rettwo{} and was not well mixed.  
The decoupling of the light-$\alpha$ and r-process enrichments is consistent with observations of Milky Way halo r-II stars  \citep{Roederer2014b, Hansen2016}.

High \xfe{Mg} ratios are a genuine characteristic of some stars in the UFD galaxies. Four of the Mg-enhanced stars (in Boo I, Seg 1, Psc II, and Car III) are also C-enhanced \citep{Norris2010, Gilmore2013, Spite2018, Ji2020}; however, they show no signs of r-process or s-process enrichments, thus they are identified as CEMP-no stars.

\begin{figure}
  \centering
  \includegraphics[scale=0.35,trim = 0.in 0.in 0.in 0.in, clip]{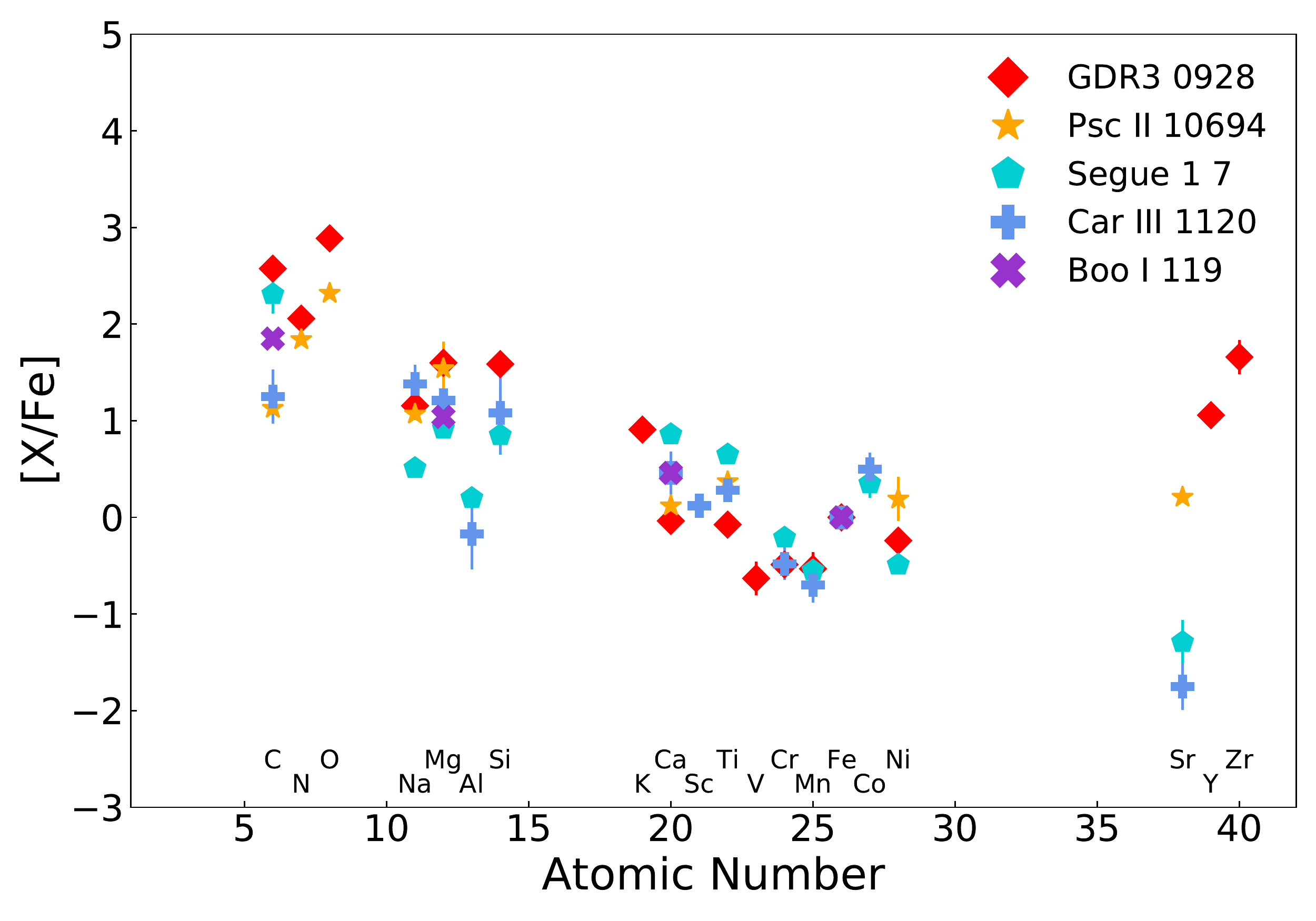}
  \caption{Chemical abundance patterns scaled to Fe for CEMP-no stars with known light-element enhancements in UFDs.  These stars cover a range of metallicities with Car \textsc{III} 1120 being the most metal-poor at \feh{} $=$ -3.9, Segue 1 7 at \feh{} $= -3.6$, Boo \textsc{I} 119 at \feh{} $= -3.3$, and \psctwo{} 10694 at \feh{} $ = -2.6$,  similar to GDR3 0928 at \feh{} $= -2.5$.}
  \label{fig:light_enrichment}
\end{figure}

If we ignore the r-process enrichment in GDR3 0928, which appears to be decoupled from the C and light $\alpha$-element enrichments, then this star would resemble the other known CEMP-no stars found in UFDs; i.e., GDR3 0928 resembles a CEMP-no star like those shown in Figure \ref{fig:light_enrichment}, with the addition of an r-II enrichment event.   

\subsection{Chemical enrichment by metal-free CCSN}

The enhanced [X/Fe] ratios for elements like C and Mg in CEMP-no stars have been interpreted as enrichment from faint SNe, where a significant fraction of the ejecta falls back onto the newly formed NS or BH \citep[e.g.,][]{Nomoto2013, Ishigaki2014, Tominaga2014, Kobayashi2020}.  While other explanations for the CEMP-no stars have been proposed (e.g., winds from rapidly rotating massive stars, accretion from AGB companions, or self-enrichment), these other sources face challenges producing a sufficient amount of elements heavier than CNO, requiring additional enrichment sources \citep[see the discussion in][]{Tominaga2014}.

Chemical enrichment modeling has shown that low-energy, faint SNe provide a natural explanation for the increased prevalence of CEMP-no stars at low metallicities ([Fe/H] $\lesssim-3$) in dwarf galaxies and the Milky Way halo, allowing dark matter halos to retain enough gas to form later populations of stars \citep{Cooke2014, Bennassuti2014}.  As a result, faint SNe could dominate the enrichment in the early universe and be a common source for the CEMP-no stars in dwarf galaxies \citep{Salvadori2015}.

Using cosmological zoom-in simulations of isolated UFDs, \citet{Jeon2021} showed that faint SNe from low-metallicity Pop II stars can contribute to the population of CEMP-no stars.  However, the stars with \xfe{C} $\gtrsim 2$ and \xfe{(Mg$+$Si)} $\gtrsim 1$ would be primarily enriched by primordial, Pop III faint SNe.  The high \xfe{C} and \xfe{Mg} ratios in GDR3 0928 favor this scenario; however, given its relatively high metallicity of \feh{} $= -2.5$ (for \rettwo{} stars), GDR3 0928 may be composed of material from both faint SNe and later generations of CCSNe \citep[e.g.,][]{Salvadori2015}.

\begin{figure}
  \centering
  \includegraphics[scale=0.32,trim = 0.in 0.in 0.in 0.in, clip]{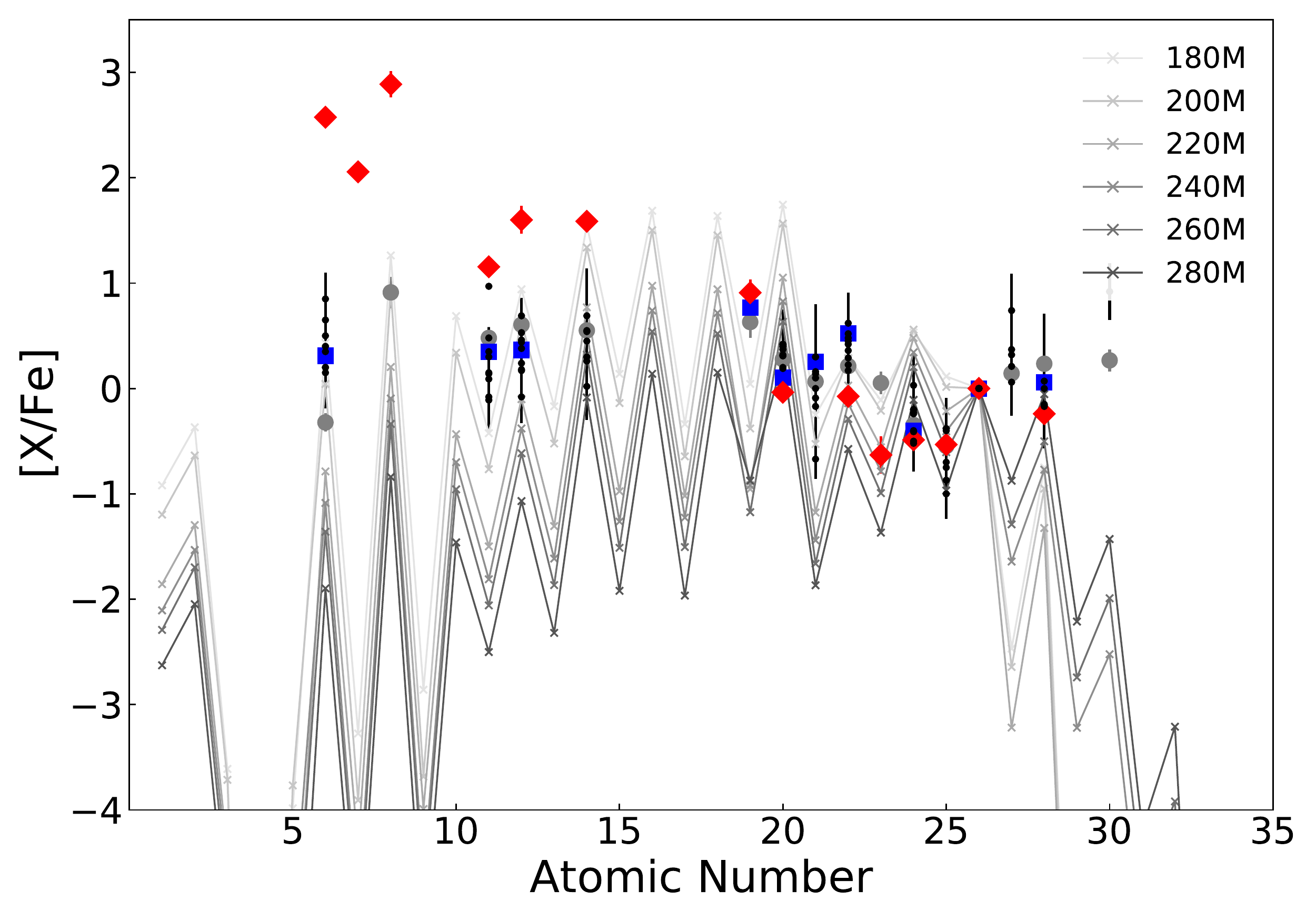}
  \includegraphics[scale=0.32,trim = 0.in 0.in 0.in 0.in, clip]{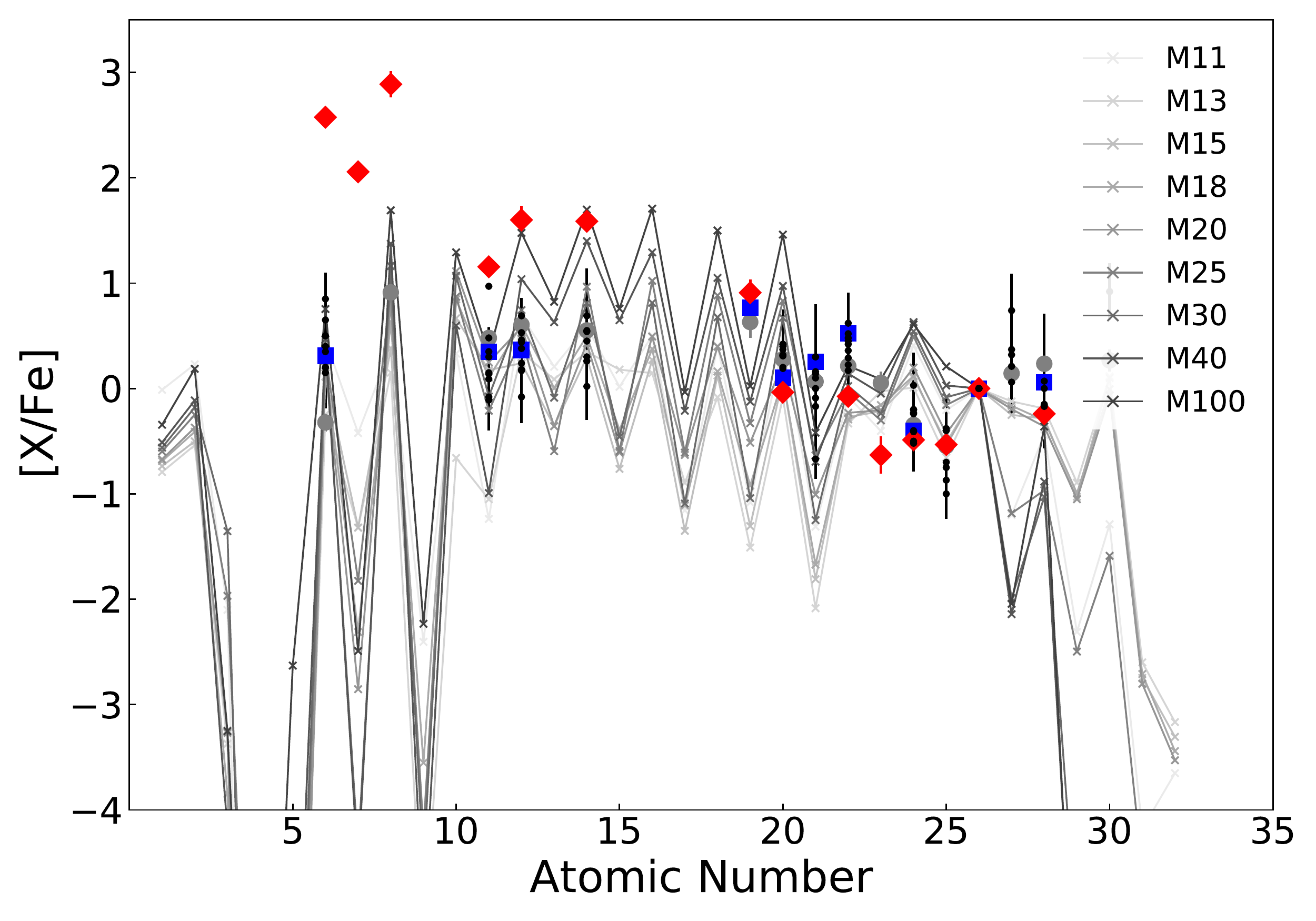}
  \caption{Comparison of the chemical abundance pattern in our two \rettwo{} stars.  (Top) Comparison to PISN yields from \citet{Takahashi2018}.  (Bottom) Comparison to metal-free CCSN (faint supernova, E51=1) yields from \citet{Nomoto2013}.
  Color-coding is the same as in Figure \ref{fig:abund_patterns}, and models of different masses are shown in grayscale (reported in solar masses).}
  \label{fig:ret2_nomoto_pisne_comp}
\end{figure}

To examine this further, we compare our abundances for GDR3 0928 to two sets of stellar yields from metal-free stars in Figure \ref{fig:ret2_nomoto_pisne_comp}.  The yields from models of Pair instability supernovae (PISN), models of metal-free massive stars with $>100$ M$_\odot$ by \citet{Takahashi2018} do not provide satisfactory fits to our stars.  
The yields from metal-free CCSN with masses $<$100 M$_\odot$ from \citet{Nomoto2013} are better, but still require additional fallback of material (i.e., faint, lower-energy metal-free CCSN) to explain the very high CNO abundances \citep{Tominaga2014}, and of course, both models require additional yields from an r-process site.

While Pop III faint SNe can be shown to produce light-element abundance enhancements like those seen in GDR3 0928, the timescales for the birth of the progenitor and subsequent death would be short, whereas \citet{Ji2023} show that efficiently mixing the r-process rich material in \rettwo{} would require timescales on the order of at least $\sim 100$ Myr.  The fact GDR3 0928's light-element enhancement is not seen in other \rettwo{} stars implies that the source of this enriched material was either a localized event (such as accretion from a companion) or not well-mixed, and happened subsequently to the r-process event that has enriched GDR3 0928 and other \rettwo{} stars.

It may, therefore, be difficult from a timing perspective to produce the light-element abundance patterns of GDR3 0928 from Pop III faint SNe.  The observation of a star with CEMP-no like light-element enhancements in \rettwo{} provides a unique window into the source of the enrichment of these stars, and may be able to help constrain the timing of CEMP-no sources in the future.

\section{Conclusion}
\label{sec:conclusion}

In this work we show some of the first science results from the new GHOST spectrograph at Gemini-S, taken during its commissioning and illustrate how this instrument will be a powerful new high-resolution spectrograph for studying metal-poor stars and dwarf galaxies.  In addition to observing the metal-poor benchmark giant, HD 122563, and a known \rettwo{} member, GDR3 1760, We have observed a new member of \rettwo{}, GDR3 0928, and find that it is a chemically peculiar star with an enhancement in light-element abundances (C, N, O, Na, Mg, and Si) as well as r-process enhancements as seen in many other \rettwo{} stars.

These light-element enhancements are thought to be independent of the r-process enhancements and resemble those that are seen in CEMP-no stars.  The light element enhancements of CEMP-no stars are commonly thought to trace the yields from the first stellar populations that produce faint supernovae with high amounts of fallback onto the stellar remnant.  Searching for more stars that share these chemical abundance patterns can also provide estimates of the total yields from these faint SNe and place limits on how well this material was mixed in its host's ISM.

These observations demonstrate the potential the new GHOST spectrograph as a powerful tool to study the chemistry of metal-poor stars in Milky Way's halo and its dwarf galaxy satellites.

\begin{acknowledgements}

We would like to thank the Gemini staff for their help and support in making the GHOST commissioning such a success.  CRH thanks Terese Hansen, Keith Hawkins, and Alex Ji, for helpful conversations about the sources of CEMP-no stars and how these observations may fit into the growing picture for these stars.

This research used the Canadian Advanced Network For Astronomy Research (CANFAR) operated in partnership by the Canadian Astronomy Data Centre and The Digital Research Alliance of Canada with support from the National Research Council of Canada the Canadian Space Agency, CANARIE and the Canadian Foundation for Innovation. 

Based on observations obtained at the international Gemini Observatory, a program of NSF’s NOIRLab, which is managed by the Association of Universities for Research in Astronomy (AURA) under a cooperative agreement with the National Science Foundation on behalf of the Gemini Observatory partnership: the National Science Foundation (United States), National Research Council (Canada), Agencia Nacional de Investigaci\'{o}n y Desarrollo (Chile), Ministerio de Ciencia, Tecnolog\'{i}a e Innovaci\'{o}n (Argentina), Minist\'{e}rio da Ci\^{e}ncia, Tecnologia, Inova\c{c}\~{o}es e Comunica\c{c}\~{o}es (Brazil), and Korea Astronomy and Space Science Institute (Republic of Korea).

This research made use of {\texttt{topcat}} \citep{topcat}, Astropy, a community-developed core Python package for Astronomy \citep{astropy}, NASA's Astrophysics Data System, and the SIMBAD database, operated at CDS, Strasbourg, France. 

\facilities{Gemini-S (GHOST spectrograph)}

\software{TOPCAT \citep{topcat}, Astropy \citep{astropy}, Matplotlib (\url{http://dx.doi.org/10.1109/MCSE.2007.55}), Numpy (\url{http://scitation.aip.org/content/aip/journal/cise/13/2/10.1109/MCSE.2011.37})}

\end{acknowledgements}

\bibliographystyle{aasjournal}
\bibliography{references}

\appendix

\section{Details of the Analysis Methodology}
\label{app:spectral_analysis}

To measure individual lines we first synthesize spectra in a range of $\pm 5$ \AA{} around the line of interest (plus a short buffer for convolving the spectra) for most atomic lines, with larger ranges for some of the more extended molecular features.  We vary the element in question from -0.6 to 0.6 in steps of 0.3 dex around a starting guess.  We then interpolate these spectra to the pixels of our observed spectra.  Using routines from the Brussels Automatic Code for Characterizing High accUracy Spectra \citep[BACCHUS][]{bacchus}, we renormalize the observed spectra and we calculate a narrower window around the line of interest from which to measure our abundances (i.e., selecting pixels around the line of interest that are sensitive to changes in the abundance of that element).

We then have two methods for measuring the abundance of each line: 1) by measuring the equivalent width of the observed line and comparing it to the equivalent widths of that line in the syntheses, and 2) measuring the abundance that minimizes the $\chi^2$ differences between the observed and synthetic spectra.  

For the equivalent width measurements, we directly integrate the observed flux within the measurement window and repeat this measurement for each of the synthetic spectra.  We then interpolate the synthetic equivalent widths to the observed equivalent width in order to determine the optimal abundance of our observed line.  We can also use the summed variance within the measurement window to estimate the uncertainty on the observed equivalent width and convert this to an abundance uncertainty.  Moreover, using these equivalent width uncertainties we can both require that our equivalent width is a $5\sigma$ detection and derive a $5\sigma$ upper limit on the abundances.

To determine our ``$\chi^2$ abundances'' we calculate the $\chi^2$ difference between each of our synthetic spectra and the observed spectrum.  We then interpolate these to identify the minimum $\chi^2$ and corresponding abundance.  If the minimum $\chi^2$ abundance is within one step of the edge of our synthesis range, we reiterate our process with a new set of synthetic spectra with a range of abundances around the minimum $\chi^2$ abundance from the initial fit.  We iterate this up to five times after which, if the abundance still lies within half a step of the edge of the synthesis range we flag the line as suspicious due to hitting the synthesis limits.

Our analysis will automatically flag lines that have hit the synthesis limits or are flagged as upper limits according to our equivalent width analysis, but we also apply a couple of other quality restrictions to our line measurements.  We also remove any lines whose line depth is less than 5$\sigma$ from the continuum.  And, to avoid using lines that are saturated to the point that they are relatively insensitive to the abundance of that element (but not so abundant as to be collisionally broadened) and would otherwise have very large uncertainties, we also remove lines where the noise in the observed spectrum is sufficiently larger than the flux differences across our syntheses.We apply these cuts automatically and do not report abundance measurements for any lines flagged in this way, but we have also visually inspected individual line fits and removed any poorly fit lines, those that are affected by badly fit blends, etc., as a final quality assurance check.

\end{document}